\documentclass{emulateapj}
\usepackage{apjfonts}
\usepackage{graphicx}
\usepackage{amsmath}
\usepackage{dpfloat}
\usepackage{wasysym}

\usepackage{longtable}

\usepackage[toc,page]{appendix}

\usepackage{color}

\def\kms{\ifmmode{\rm km\thinspace s^{-1}}\else km\thinspace s$^{-1}$\fi}

\shortauthors{Rappaport et al.~2016}
\shorttitle{Drifting Asteroids}

\begin{document}

% ------------------------------------------------------------------------
% New commands
%
\def\ltsima{$\; \buildrel < \over \sim \;$}
\def\lsim{\lower.5ex\hbox{\ltsima}}
\def\gtsima{$\; \buildrel > \over \sim \;$}
\def\gsim{\lower.5ex\hbox{\gtsima}}
\def\teff{$T_\mathrm{eff}$}                 
\def\vsini{\hbox{$v$\,sin\,$i_{\star}$}}    
\def\cms2{\hbox{\,cm\,s$^{-2}$}}

% -------------------------------------------------------------------------
%

\def\kms{\ifmmode{\rm km\thinspace s^{-1}}\else km\thinspace s$^{-1}$\fi}

\bibliographystyle{apj}

\title{Drifting Asteroid Fragments Around WD 1145+017}

\author{
S.~Rappaport\altaffilmark{1}, 
B.L.~ Gary \altaffilmark{2},
T.~ Kaye \altaffilmark{3},
A.~Vanderburg\altaffilmark{4},
B.~ Croll \altaffilmark{5},
P.~ Benni \altaffilmark{6},
J.~ Foote \altaffilmark{7},
}

\altaffiltext{1}{Department of Physics, and Kavli Institute for Astrophysics and Space Research, Massachusetts Institute of
  Technology, Cambridge, MA 02139, USA; sar@mit.edu}
\altaffiltext{2}{Hereford Arizona Observatory, Hereford, AZ 85615; BLGary@umich.edu}
\altaffiltext{3}{Raemor Vista Observatory, 7023 E. Alhambra Dr., Sierra Vista, AZ 85650; tom@TomKaye.com}
\altaffiltext{4}{Harvard-Smithsonian Center for Astrophysics, 60 Garden
Street, Cambridge, MA 02138 USA}
\altaffiltext{5}{Institute for Astrophysical Research, Boston University, 725
Commonwealth Ave. Room 506, Boston, MA 02215; croll@bu.edu}
\altaffiltext{6}{Acton Sky Portal, 3 Concetta Circle, Acton, MA 01720; pbenni@verizon.net}
\altaffiltext{7}{Vermillion Cliffs Observatory, 4175 E. Red Cliffs Dr., Kanab, UT 84741; jfoote@kanab.net}

\slugcomment{Submitted to {\it Monthly Notices of the Royal Astronomical Society}, 2016 January 27}

\begin{abstract}
We have obtained extensive photometric observations of the polluted white dwarf WD 1145+017 which has been reported to be transited by at least one, and perhaps several, large asteroids with dust emission. Observation sessions on 37 nights spanning 2015 November to  2016 January with small to modest size telescopes have detected 237 significant dips in flux.  Periodograms reveal a significant periodicity of 4.5004 hours consistent with the dominant (``A'') period  detected with K2.  The folded light curve shows an hour-long depression in flux with a mean depth of nearly 10\%.  This depression is, in turn, comprised of a series of shorter and sometimes deeper dips which would be unresolvable with K2.  We also find numerous dips in flux at other orbital phases.  Nearly all of the dips associated with this activity appear to drift systematically in phase with respect to the ``A'' period by about 2.5 minutes per day with a dispersion of $\sim$0.5 min/d, corresponding to a mean drift period of 4.4928 hours.  We are able to track $\sim$15 discrete drifting features.  The ``B''--``F'' periods found with K2 are not detected, but we would not necessarily have expected to see them.  We explain the drifting motion as due to smaller fragmented bodies that break off from the asteroid and go into a slightly smaller orbit.  In this interpretation, we can use the drift rate to determine the mass of the asteroid, which we find to be $\approx 10^{23}$ grams, or about 1/10th the mass of Ceres.  
\end{abstract}


\section{Introduction}

A growing body of evidence suggests that many white dwarfs are in the process of accreting tidally disrupted planetesimals from the progenitor stars' planetary systems. Spectroscopic detections of metal lines in 25-50\% of all white dwarfs indicate that refractory materials were recently deposited onto the surfaces of these stars (e.g., Zuckerman et al.~2010; Koester et al.~2014). The composition of the accreted material and total deposited mass can be consistent with the accretion of large asteroids like Ceres or Vesta in our own solar system (Zuckerman et al.~2007; G\"ansicke et al.~2012; Farihi et al.~2013).  Additionally, dusty debris disks have been found around many of these polluted white dwarfs (Zuckerman et al.~1987), linking the presence of elements in the photosphere to circumstellar material near the star's tidal disruption radius for rocky material (Kilic et al.~2006; Farihi et al.~2009; Barber et al.~2012; Rocchetto 2015). The accepted explanation for these observations is that after the white dwarf's progenitor leaves the main sequence and undergoes mass loss, the progenitor's planetary system destabilizes, perturbing planetary orbits close enough to the white dwarf for tidal disruption (Debes \& Sigurdsson 2002; Debes et al.~2012; Mustill et al.~2014; Veras et al.~2014; 2015).  

Recently, Vanderburg et al.~(2015) reported compelling new evidence for this scenario by detecting transits of the white dwarf WD 1145+017 (see Table \ref{tbl:wd1145} for its properties) using data from the K2 mission (Howell et al.~2014).  Vanderburg et al.~(2015) detected six distinct periodicities in K2 data ranging from 4.5 to 4.9 hours (designated as ``A'' through ``F''), and found evidence that the depth and shape of the transits varied with time. Higher cadence ground-based follow-up photometry revealed several sets of asymmetric transits with varying depths (of up to 40\%) at an orbital period of $\approx$ 4.5 hours, but with highly inconsistent transit times (Vanderburg et al.~2015). Asymmetric transit shapes and variable transit depths have come to be associated with disintegrating planets transiting main sequence stars (Rappaport 2012, 2014; Sanchis-Ojeda 2015), and Vanderburg et al.~(2015) interpreted the observations of WD 1145+017 as being caused by a similar phenomenon. Vanderburg et al.~(2015) showed that a dust cloud formed from material sublimated off several fragments of a large, tidally disrupted asteroid (with roughly the mass of Ceres) occulting the star could plausibly explain the observed transits. Finally, Vanderburg et al.~(2015) showed that WD 1145+017 shows pollution from refractory elements in its photosphere and hosts a dusty debris disk, linking the transiting disintegrating bodies to the classic picture of disrupted minor planets polluting the surfaces of white dwarfs.

Further ground-based observations of WD 1145+017 obtained in early 2015 (Croll et al.~2015) found an additional dozen transits of variable depth and  phasing that were inconsistent with any of the six periods found by Vanderburg et al.~(2015).  Croll et al.~(2015) used the times of these transits to provide the first evidence for multiple disintegrating bodies transiting WD 1145+017 with periods close to $\sim$4.5 hours, although the main period they found was shorter than the dominant period in the K2 data by about 25 seconds.  Xu et al.~(2016) presented high-resolution optical spectroscopy showing circumstellar gas absorption features near transitions of many elements detected, including some with linewidths of 300 km s$^{-1}$. These spectroscopic observations provided independent confirmation for the presence of circumstellar material around the white dwarf along our line of sight. Recently, G{\"a}nsicke et al.~(2015) presented photometric observations from late 2015 (six months after the ground-based observations of Vanderburg et al.~2015 and Croll et al.~2015) showing a qualitatively similar picture to Croll et al.~(2015), but with significantly stronger and more frequent transit features than previously seen.

While a strong case for a scenario involving disintegrating planetesimals transiting WD 1145+017 exists, this appears to be an incomplete picture of the system. In the discovery paper, Vanderburg et al.~(2015) readily admitted that they were unable to describe all of their observations with the simple model proposed for disintegrating planets around main sequence stars. Unexplained observations include: the significantly longer inferred duration of transits in K2 data compared to ground-based data; the slightly different periods detected by Vanderburg et al.~(2015) in K2 data and Croll et al.~(2015) from the ground; the non-detection by Croll et al.~(2015) of the five weaker periods detected by Vanderburg et al.~(2015) in K2 observations; and perhaps most vexing, the evidence for multiple occulting bodies in nearly the same 4.5 hour orbit to within 0.2\% in period (Croll et al.~2105; G\"ansicke et al.~2015).

In this paper, we present photometric observations of WD 1145+017, which we obtained in an attempt to understand some of these unexplained phenomena. We observed WD 1145+017 between 2015 Nov 1 and 2016 Jan 21 from four amateur observatories, comprising 53 observing sessions on 37 nights, for a total of 192 hours on target.  We detected some 237 transit events which we believe significantly help clarify our understanding of this enigmatic object.  In Sect.~\ref{sec:obs}, we describe the telescopes and instruments used in our observation and our data reduction procedures.  Our basic analyses of the dataset are presented in Sect.~\ref{sec:analysis}.  There we report on a search for periods and an analysis of the ``drifting'' behavior found for a number of the dips.  In Sect.~\ref{sec:interpret}, we present a modified physical model and interpret our results in this context.  We discuss in Sect.~\ref{sec:discuss} how the proposed model is able to explain a number of the observational results, as well as a number of its shortcomings.  Finally, in Sect.~\ref{sec:summary} we summarize our results and draw some final conclusions. 

\begin{deluxetable}{lc}
\centering

\tablecaption{Some Properties of the WD 1145+017 System}
\tablewidth{0pt}
\tablehead{
\colhead{Quantity} &
\colhead{Value\tablenotemark{a}} 
}

\startdata
RA  (J2000) &  11:48:33.59 \\ % 627  \\
Dec (J2000) &  +01:28:59.3 \\ % 41 \\
Distance &  $174 \pm 25$ pc \\
Mass &  $0.6 \pm 0.05 ~ M_\odot$\\
Radius &  $1.34 \pm 0.14 ~ R_\oplus$ \\ 
$T_{\rm eff}$ &  $15,900 \pm 500$ K \\
$L_{\rm bol}$ &  $0.0088 \pm 0.0021$ \\
V-magnitude & $17.24 \pm 0.02$ \\
B-V & $-0.08 \pm 0.04 $ \\
$P_{\rm A}$\tablenotemark{b} &  $4.4989 \pm 0.0001$ hr\\
$a_{\rm A}$\tablenotemark{c} &  0.0054 AU \\
$a_{\rm A}$\tablenotemark{c} &  94 ~ $R_{\rm wd}$ \\ 
$a_{\rm A}$\tablenotemark{c} &  1.16 ~ $R_\odot$ \\
Asteroid radius\tablenotemark{d} & $\approx 200 ~{\rm km} \simeq 12\% \,R_{\leftmoon}$ \\
\enddata
\tablenotetext{a}{Adapted from Vanderburg et al.~(2015).}
\tablenotetext{b}{Orbital period of the ``A'' asteroid (Vanderburg et al.~2015).}
\tablenotetext{c}{Semimajor axis of the ``A'' asteroid; see text.}
\tablenotetext{d}{Based on a mass of $10^{23}$ g, inferred from the model presented in this work (Sect.~\ref{sec:interpret}), and a mean density of 3 g/cc.}
\label{tbl:wd1145}
\end{deluxetable}

\section{Observations}
\label{sec:obs}   

\subsection{Observing Facilities}

All of the observations were carried out at a set of privately operated amateur observatories.  The telescopes range in aperture from 28 cm to 80 cm.  The observatories, their locations, and their equipment are summarized in Table \ref{tbl:facilities}.  Here we list the observers and briefly describe some of the details associated with their observations.

Bruce Gary uses a 35-cm fork-mounted Meade Schmidt-Cassegrain (``S-C'') telescope at the Hereford Arizona Observatory (HAO) in Hereford, Arizona, with Minor Planets Center (MPC) observatory code G95. The telescope, off-axis autoguider, SBIG ST-10XME CCD, and dome are controlled via buried cables using the MaxIm DL 5.2 control program. 

Tom Kaye, of the Raemor Vista Observatory, uses an 80-cm Ritchie-Chretien telescope in a 16-foot dome located near Sierra Vista, AZ. An Apogee U47, 1k $\times$ 1k back-illuminated, E2V chip CCD is used for remotely-controlled observing. Dedicated autoguiding with a 2nd CCD is not performed, but in its place plate-solving of each 15-second calibrated exposure is followed by pointing nudges, if necessary. The observatory continues to be referred to by its original name, `Junk Bond Observatory' (JBO), given to it by the original owner, Dave Healy (deceased).

Paul Benni operates several telescopes at the Acton Sky Portal private observatory, in Acton, Mass. Celestron 28-cm and 35-cm Schmidt-Cassegrain telescopes, employ off-axis autoguiders on Losmandy mounts. Imaging with SBIG ST-8XME CCDs is performed using the TheSkyX program. 

Jerry Foote operates telescopes at the Vermillion Cliffs Observatory with MPC observing code G85. For these observations a 40-cm f/3.5 prime focus fork mounted telescope with an SBIG ST-7 CCD behind a coma corrector was used with off axis guiding. 

A log of all the observations of WD 1145+017 made at these facilities is given in Table \ref{tbl:log}.
 
\begin{deluxetable*}{lcccc}
\centering
\tablecaption{Observing Facilities}
\tablewidth{0pt}
\tablehead{
\colhead{} &
\colhead{Observatory~1} &
\colhead{Observatory~2} &
\colhead{Observatory~3} &
\colhead{Observatory~4}
}
\startdata
Observatory Name & Hereford Arizona Observatory &  Raemor Vista Observatory  & Acton Sky Portal & Vermillion Cliffs Observatory \\
Observer & Bruce Gary & Tom Kaye & Paul Benni & Jerry Foote \\
Telescope & Meade 35-cm S-C  & 80-cm Ritchie-Chretien & Celestron 28-cm and 35-cm S-C  &  40-cm S-C\\
Latitude &	+31.45  & +31.48	 & +42.45  & +37.03  \\
E-Longitude & $-$110.24 & $-$110.20 & $-$71.43  & $-$112.43  \\
Filter & clear	 &  clear	& clear	 & clear	 \\
Exposure & 40 s & 15 s	&  40 s	 & 60s	 \\
Number of reference stars & $\approx$18 & $\approx$18 & 5 & 1 \\
\enddata
\label{tbl:facilities}
\end{deluxetable*}

\subsection{Data Reduction}

Most of the observations reported here are by Gary (35 cm) and Kaye (80 cm).  The images have original plate scales of 0.72$''$ and 0.61$''$ per pixel, respectively, and are then binned into $2 \times 2$ pixels.  Because the WD 1145+017 observations began 4.5 months before the most favorable observing season, all observations were initially made at low elevations, where seeing was poor and the point-spread-function (`PSF') of star images was large. The PSF underwent large changes during each observing session, so aperture photometry (after bias, dark, and twilight flat calibrations were applied) could not be performed with a single set of photometric apertures.  Instead, images were grouped by air mass and circular photometry radii were chosen to be a multiple (typically 2.0) of a typical PSF FWHM.  For example, when the FWHM was 3.0$''$, a circular photometry radius corresponding to 6.0$''$ was typically used for the photometric aperture, and therefore multiple aperture radii were often used to process the data for each observing session. For determining the sky background we typically employed an annulus width and gap width of 15$''$ each.  A selection of photometric aperture sizes was evaluated for their effect on the light curve noise levels and estimated systematic errors, and a subjective choice was adopted that was appropriate for each air-mass segment of data.

The numbers of reference stars chosen for photometry measurement were within the range of 19 to 23, depending on the stability of the field-of-view (e.g., tracking shortcomings occasionally caused some stars, meant for use as reference, to be outside the FOV).  At a later stage of analysis we evaluated the contribution of each reference star to the light curve noise level, and a subset of typically 18 reference stars was adopted for final use. Since much of the data were taken at high air mass, the effect of red stars for calibrating a blue target star was a serious concern, and this situation was avoided.

The uncertainty of the WD 1145+017 (target star) magnitude measurements was estimated with a ``neighbor difference'' method that we describe below (Gary 2014). Nearby stars that were slightly brighter than the target were analyzed, and a standard error (`SE') vs.~magnitude model was used to predict the SE for each target star measurement.  The neighbor difference method begins with squaring the difference of a star's magnitude with the average of the same star's temporal neighbor magnitudes, dividing by $\sqrt{1.5}$, and forming a running average for a couple dozen such values, after which the square-root is an estimate of that star's SE.  

On nights when observations were made by two or three observers we applied magnitude (vertical) offsets to the data so that the out-of-transit (OOT) portions of the light curves matched. Most of the $\sim$18 reference stars had APASS magnitudes (g$'$ to i$'$), and these were used to determine that WD 1145+017 had r$'$-mag = 17.22 during OOT.  A method equivalent to the standard ``CCD transformation equations'' was used, on the assumption of an adopted target star color (g$'$- r$' = -0.20$). 

Whereas the Gary and Kaye observations were processed using the MaxIm DL program, which is popular with amateur observers, the observations by Benni (28 cm and 35 cm) and Foote (40 cm) were processed using AIP4WIN, another program in common use by amateur observers. The numbers of reference stars for these observations were 5 and 1, respectively.   Aperture photometry was performed and data files with Julian Date time tags, magnitudes, and SEs were treated in a way similar to the above description.  For these data the neighbor difference method was used to estimate SEs for each target magnitude using target magnitudes, not nearby reference stars. This will overestimate the SEs when the target undergoes abrupt changes, which is one shortcoming for these data sets.
 
A summary plot of all WD 1145+017 light curves is presented in Fig.~\ref{fig:data}.  Each night of data is stacked vertically below the next, with no spaces left for missing nights.  The data are all phased with respect to the ``A'' period of 4.5004 hours (see Sect.~\ref{sec:analysis} for definition).  Significant dips are noticeable on virtually all nights where we used telescope apertures of 35 cm or greater.  On some nights as many as 13 dip events are observed, with dip depths that can reach 60\%.  Many, but certainly not all, of the dips tend to congregate within $\pm$ 1/2 hour around phase 0 (see Fig.~\ref{fig:data} and Table \ref{tbl:periods} for definitions); however, dips are also observed at numerous other phases about the orbit.

\section{Analysis}
\label{sec:analysis}

\subsection{Period Search}
\label{sec:periods}

In order to search for frequencies in our data set, we employed a number of standard algorithms, including a simple FFT, a Lomb-Scargle transform with and without the inclusion of uncertainties on the individual data points, a Stellingwerf periodogram (Stellingwerf 1978), and a box least-squares transform (`BLS'; Kov\'acs et al.~2002).  In the end, we chose the BLS transform as the most robust for this data set, and for its ability to track narrow dipping features.  Somewhat surprisingly, the BLS was actually less susceptible to the 1-day observing window function than was the Lomb-Scargle transform.

\begin{figure}[h]
\begin{center}
\includegraphics[width=0.476 \textwidth]{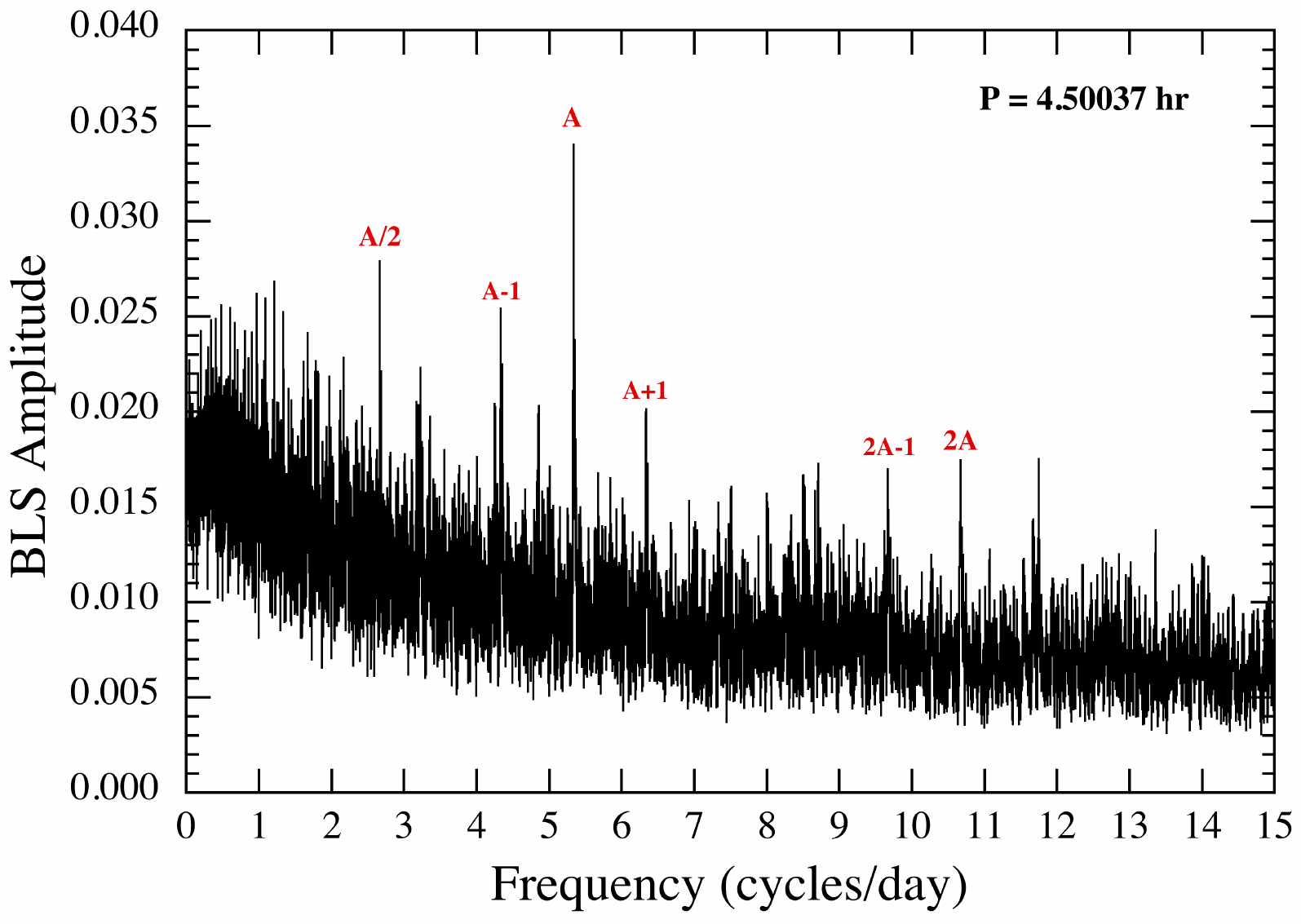} 
\caption{Box least squares transform of the entire data set.  The more prominent peaks are labeled with the letter ``A'' in reference to the K2 period with the largest amplitude.  The ``2'' preceding the ``A'' denotes whether the peak belongs to the first harmonic of the base frequency. The number following the ``A'' denotes the particular sideband of the 1-day observing window.  The best period for ``A'' determined from this BLS-transform is 4.5004 $\pm 0.0013$ days, or within 3 parts in $10^4$ of the ``A'' period measured with K2.  }
\label{fig:LS1}
\end{center}
\end{figure}

The BLS transform of the entire data set is shown in Fig.~\ref{fig:LS1}.  We have labeled the more prominent peaks ``A'' in reference to the K2 period with the largest amplitude (Vanderburg et al.~2015); the ``2'' preceding the ``A'' denotes that the peak belongs to the next harmonic after the base frequency. The number following the ``A'' denotes the particular sideband of the 1-day observing window.  Essentially all of the significant peaks in the BLS transform result from the ``A'' period or its higher harmonics and sidebands of the 1-day observing window.  The best period for ``A'', determined from this BLS transform is 4.5004 $\pm 0.0013$ hours, which agrees with the K2 ``A'' period to three parts in $10^4$, and to within their mutual uncertainties. 

We interpret this to indicate that the ``A'' period first observed with K2 is still persistent and stable $\sim$1.5 years later.  This will be important for our interpretation of the observations.

\begin{deluxetable}{lcccc}
\centering
\tablecaption{WD 1145+017 Observation Log}
\tablewidth{0pt}
\tablehead{
\colhead{UTC Date} &
\colhead{Duration (hr)} &
\colhead{Aperture} &
\colhead{Observer} &
\colhead{No. Dips}
}
\startdata
Nov 01 & 1.3 & 14$''$ & Gary &	0 \\
Nov 08 &	1.4 & 14$''$ & Gary  & 1 \\
Nov 11 &	1.5 &	14$''$  & Gary  & 0 \\
Nov 15 &	0.8 & 11$''$ & Benni 	&	1 \\
Nov 18 &	2.7 & 14$''$ & Gary	 &	2 \\
Nov 19 &	2.9 & 14$''$ & Gary	 & 	4 \\
Nov 21 &	3.0 & 14$''$ & Gary	 &	7 \\
Nov 22 &	2.9 & 14$''$ & Gary	 &	5 \\
Nov 24 &	1.9 & 16$''$ & Foote  	&	  4 \\
Nov 24 &	1.8 & 11$''$ & Benni 	&	  3 \\
Nov 24 &	2.9 & 32$''$ & Kaye	 & 	4 \\
Nov 25 &	1.8 & 11$''$ & Benni 	&	4 \\
Dec 02 &	2.9 & 32$''$ & Kaye 	&	4 \\
Dec 03 &	3.0 & 32$''$ & Kaye	 &	5 \\
Dec 04 &	2.1 & 11$''$ & Benni 	& 3 \\
Dec 05 &	2.5 & 11$''$ & Benni 	&	2 \\
Dec 06 &	2.6 & 11$''$ & Benni 	&	0 \\
Dec 06 &	4.3 & 14$''$ & Gary	 & 	6 \\
Dec 09 &	4.4 & 32$''$ & Kaye	 &	6 \\
Dec 14 &	4.0 & 14$''$ & Gary	 &	5 \\
Dec 14 &	4.0 & 32$''$ & Kaye 	&	4 \\
Dec 16 &	4.4 & 14$''$ & Gary	 &	6 \\
Dec 16 &	4.1 & 32$''$ & Kaye 	& 	12 \\
Dec 18 &	3.9 & 14$''$ & Gary	 & 	6 \\
Dec 18 &	3.5 & 32$''$ & Kaye	 &	3 \\
Dec 20 &	3.7 & 11$''$ & Benni 	&	3 \\
Dec 21 &	4.9 & 14$''$ & Gary	 &	3 \\
Dec 21 &	4.6 & 32$''$ & Kaye 	&	8 \\
Dec 28 &	2.5 & 14$''$ & Gary	 & 	4 \\
Dec 29 &	3.6 & 14$''$ & Gary 	&	3 \\
Dec 29 &	3.6 & 32$''$ & Kaye	 &	7 \\
Dec 30 &	4.2 & 32$''$ & Kaye	 &	12 \\
Jan 02 &	4.5 & 32$''$ & Kaye 	&	5 \\
Jan 03 &	5.2 & 32$''$ & Kaye	 &	9 \\
Jan 04 &	2.0 & 14$''$ & Benni 	&	1 \\
Jan 05 &	4.7 & 14$''$ & Benni 	&	6 \\
Jan 06 &	4.5 & 14$''$ &	Benni & 6 \\
Jan 07 &	5.0 &	 14$''$ &	Benni & 5 \\
Jan 10 &	6.9 &	 14$''$ &	Gary	& 10 \\
Jan 10 & 	6.0 &	 32$''$ &	Kaye	 & 13 \\
Jan 11 &	5.9 & 32$''$ & Kaye & 10 \\
Jan 12 &	6.8 & 14$''$ & Gary & 9 \\
Jan 12 &	6.1 & 32$''$ & Kaye	& 9 \\
Jan 12 &	6.5 & 16$''$ & Foote	& 4 \\
Jan 12 &	2.2 & 14$''$ & Benni	 & 3 \\
Jan 13 &	3.0 & 32$''$ & Kaye	& 9 \\
Jan 13 &	6.7 & 16$''$ & Foote	& 9 \\
Jan 18 & 6.8 & 14$''$ & Gary & 10 \\
Jan 20 & 3.0 & 14$''$ & Benni & 3 \\
Jan 21 & 1.4 & 14$''$ & Gary & 1 \\
Jan 21 & 4.9 & 32$''$ & Kaye & 5 \\
\enddata
\label{tbl:log}
\end{deluxetable}

\begin{figure*}
\begin{center}
\includegraphics[width=0.7 \textwidth]{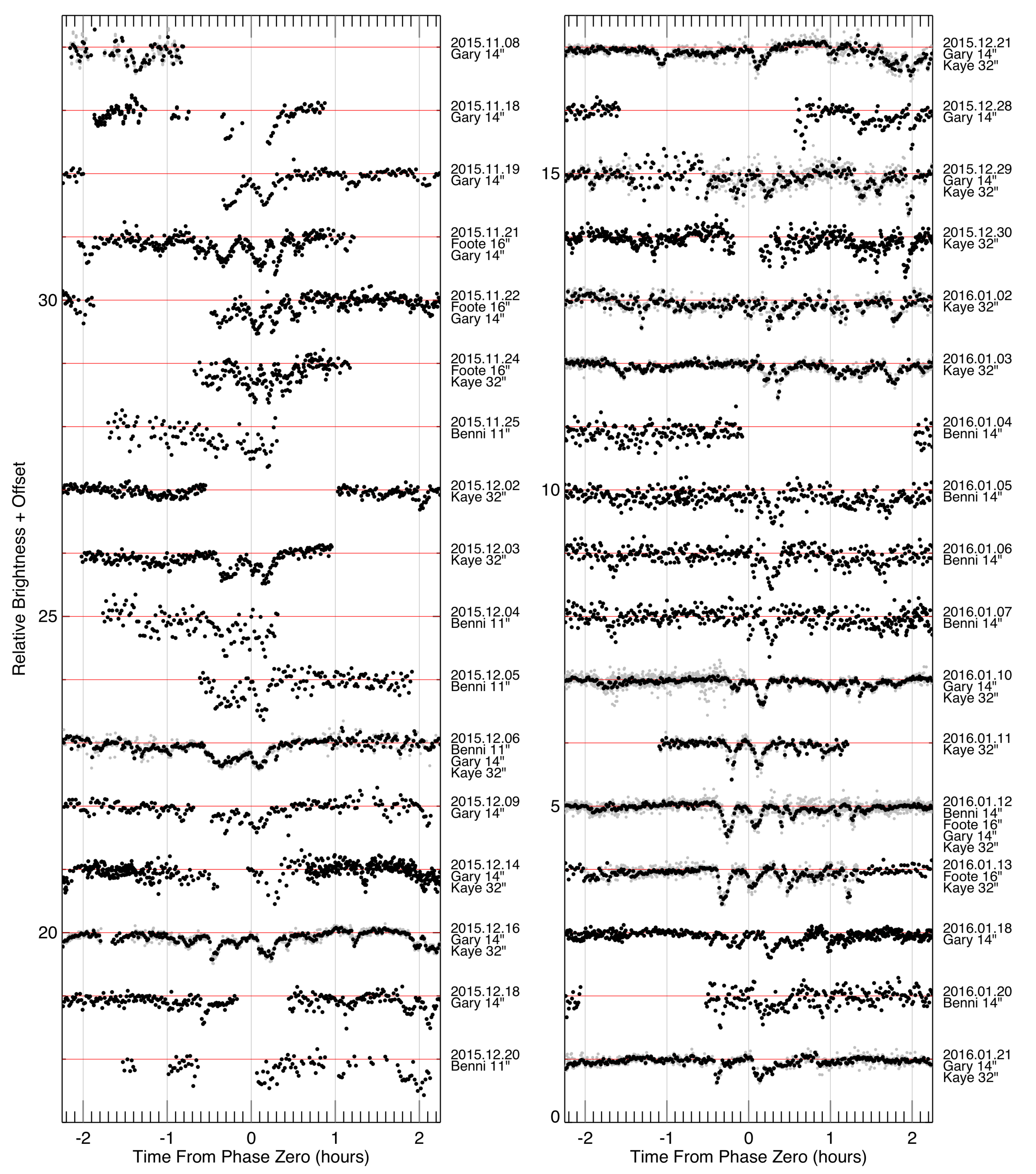}
\caption{A summary plot of all the flux data acquired  from WD 1145+017 for this work during 2015 November and December and 2016 January showing the night-to-night variations in the dipping behavior.  The grey points are individual flux measurements while the darker points are 1-minute averages.  The data are all phased to a common period of 4.5004 hours (see Table \ref{tbl:periods}), and the BJD corresponding to zero phase is given in Table \ref{tbl:periods}.  The observation date (UTC) and the telescope are indicated next to each panel.  A log of the observations is given in Table \ref{tbl:log}. }
\label{fig:data}
\end{center}
\end{figure*}

\begin{figure}
\begin{center}
\includegraphics[width=0.48 \textwidth]{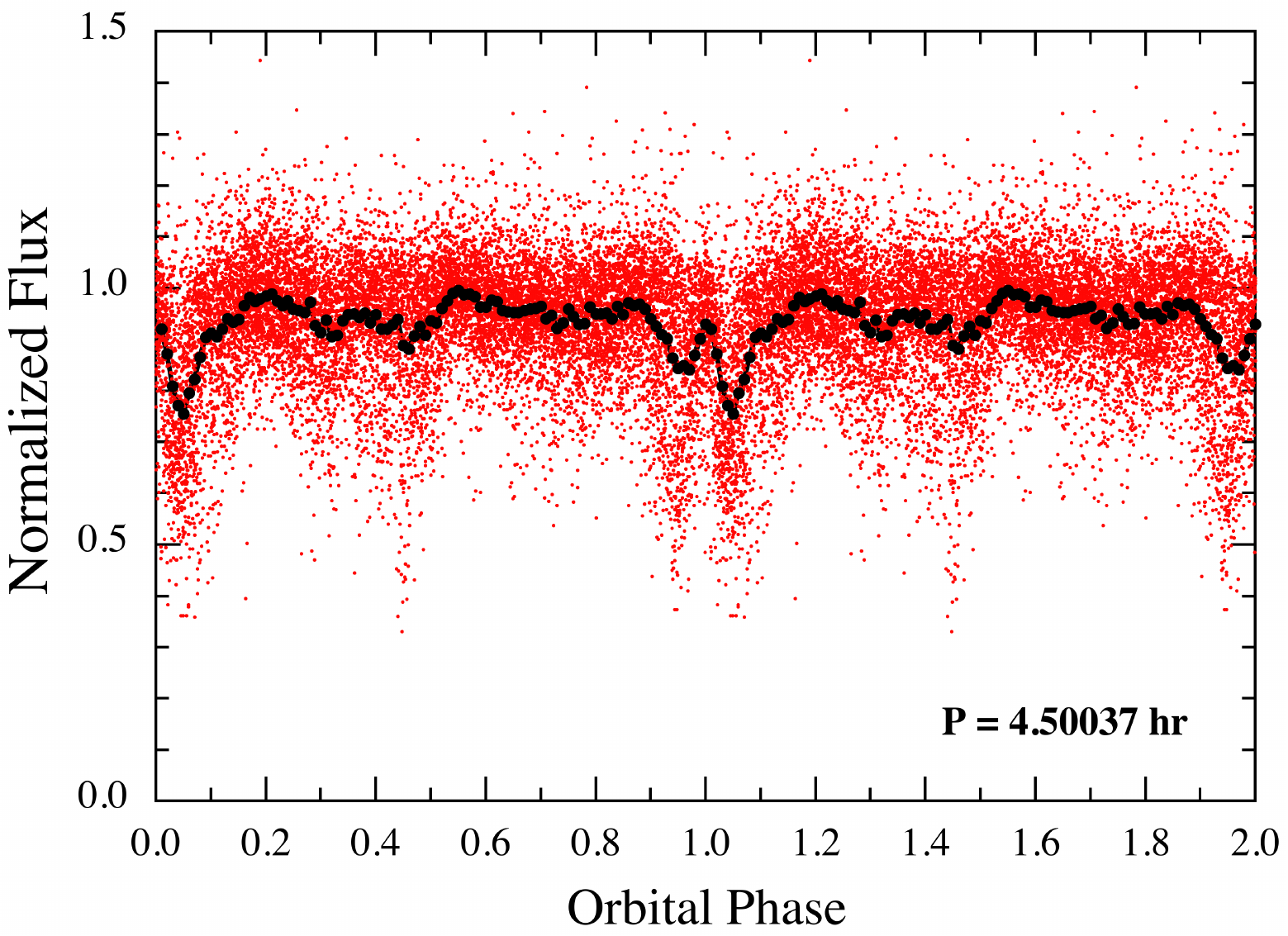}
\caption{A fold of all the flux data modulo a period of 4.5004 hours. The epoch of phase zero is defined in Table \ref{tbl:periods}.  This phase zero is shifted by about a half a cycle from the K2 phase zero 1.5 years earlier, and is just within the uncertainties of the projection (see discussion in Sect.~\ref{sec:periods}).  Red points are the individual flux measurements.  The data in each of 100 phase intervals are averaged by weighting  according to the uncertainties in the individual flux measurements.  These binned averages are plotted as black symbols.  Phase 0 is defined in Table \ref{tbl:periods}.
}
\label{fig:fold1}
\end{center}
\end{figure}

A fold of the data about a period of 4.5004 hours for the entire data set is shown in Fig.~\ref{fig:fold1}.  The fold shows all the individual flux points so that the variations are evident, as well as a binned and averaged version of the lightcurve (in 100 bins).  The binned fold uses the relative uncertainties in the individual flux measurements (see Sect.~\ref{sec:obs}) in order to weight the points within a given bin.  The dominant feature is an hour-long depression near phase zero.   

We define phase zero as the midpoint between the two closely spaced dips in Fig.~\ref{fig:fold1}, which occurs at BJD = 2457316.686 with a period of  4.5004 hours.  

We note that technically there is a statistically significant offset between our definition of phase zero for our ``A'' period, and the phase zero for the ``A'' period as defined in Vanderburg et al.~(2015) from the K2 observations. 
The ``A'' period determined from the K2 observation, coupled with the K2 phase zero, projects ahead to our ground-based observations such that it is $\approx 1/2$ cycle off our phase zero (defined above) with an uncertainty of $\approx \pm \, 1/20$ cycle.  This is based on the formal uncertainty of the K2 ``A'' period given in Table \ref{tbl:periods} which, in turn, is inferred from an uncertainty of $\approx 3$ minutes in the determination of the times of minima.  This latter uncertainty is justified by the statistical precision of the K2 signal.  However, due to the limited time resolution of K2 (see below) these observations were insensitive to the type of systematic and erratic behavior of the dips exhibited in Fig.~\ref{fig:data}.  In light of this new understanding of the dip behavior, it seems reasonable that the uncertainty in the K2 ``A'' period (Vanderburg et al. 2015) is actually an order of magnitude larger than cited in Tables \ref{tbl:wd1145} and  \ref{tbl:periods}.  Thus, we cannot uniquely infer the number of orbits over the $\approx 2700$ cycles between the two sets of observations.  Therefore, it seems prudent to use the phase zero determined by the current observations as defined above and reported in Table \ref{tbl:periods}.

\begin{figure}
\begin{center}
\includegraphics[width=0.48 \textwidth]{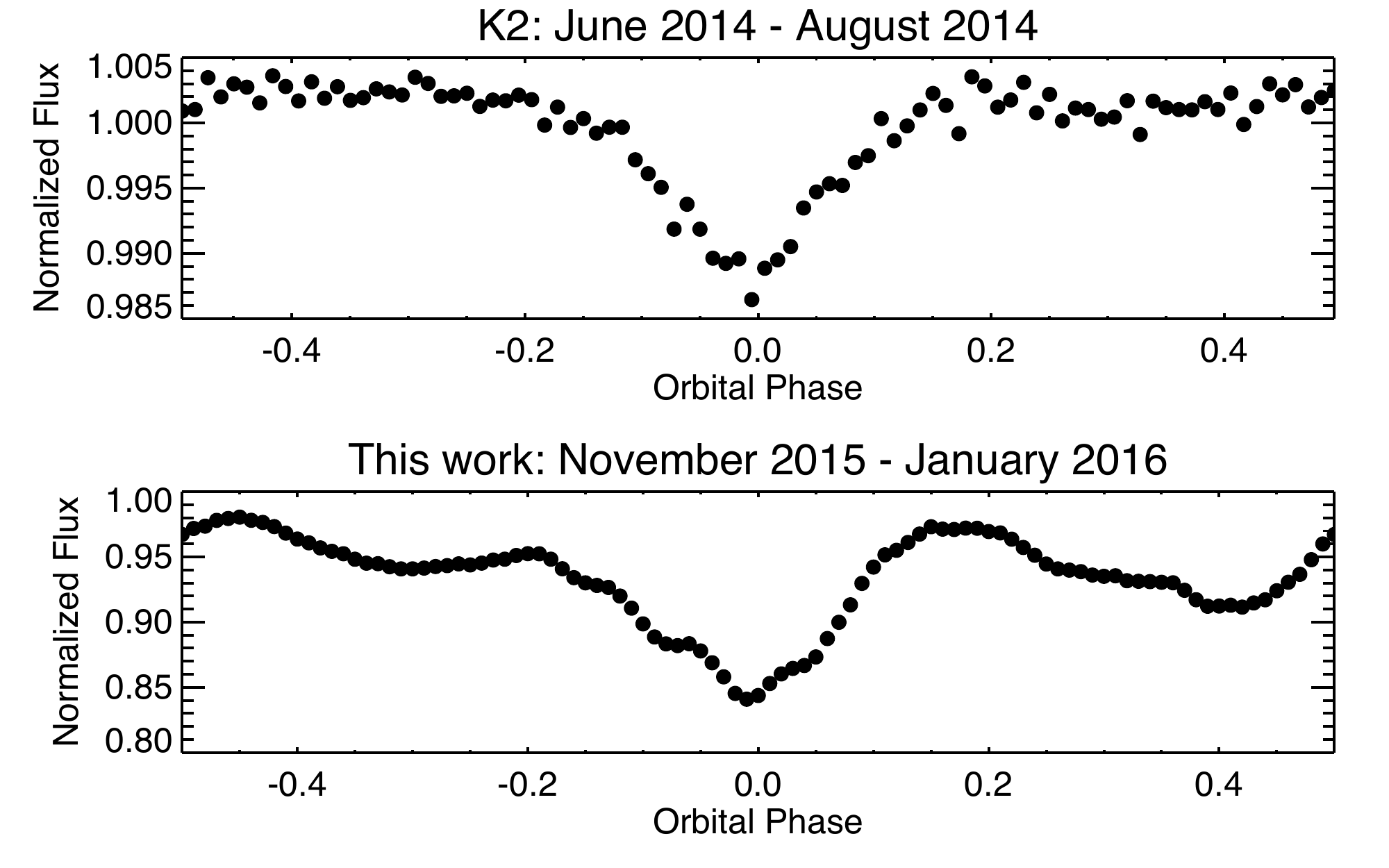}
\caption{Comparison of the K2 data (top panel) folded about the ``A'' period (from Vanderburg et al.~2015) vs.~the data from the current ground-based observations (bottom panel) folded about the ``A'' period but also convolved with the 29.4 min long-cadence integration time of K2.  Each panel is referenced to its own phase zero defined in Table \ref{tbl:periods}. This puts the two observations on a more nearly equal footing, and shows how deep the mean flux depression is currently as compared to what it was during the K2 epoch.  It is important to note that the flux scale on the bottom panel is an order of magnitude larger than on the top panel.  
}
\label{fig:convolve}
\end{center}
\end{figure}

We have convolved the folded light curve in Fig.\ref{fig:fold1} with the 29.4 min long cadence integration time of {\em Kepler} to simulate what our ground-based data would have looked like if observed with K2 (see Fig.~\ref{fig:convolve}).  This yields a rather similar profile to that seen with K2 as reported by Vanderburg et al.~(2015), including an hour-long depression in flux (Fig.~\ref{fig:convolve}).   For example, the all-data fold in Fig.~\ref{fig:fold1} would be seen by {\em Kepler} as a 10\% deep dip centered on phase zero. During the $\sim$2.5 months of the observations reported here, that dip depth varies from $\sim$25\% to 10\% and back to 20\%. This contrasts with the K2 depths that ranged from nearly 0\%, to 2\%, and to 1\% over a similar $\sim$2.6 month interval; in other words, both the K2 and present data exhibit changes in gross activity level, but since the K2 data of $\sim$1.5 years ago, the level of ``activity'' of the transits has increased by about an order of magnitude.

\subsection{Analysis of Drifts}
\label{sec:drifts}

\begin{figure}[h]
\begin{center}
\includegraphics[width=0.48 \textwidth]{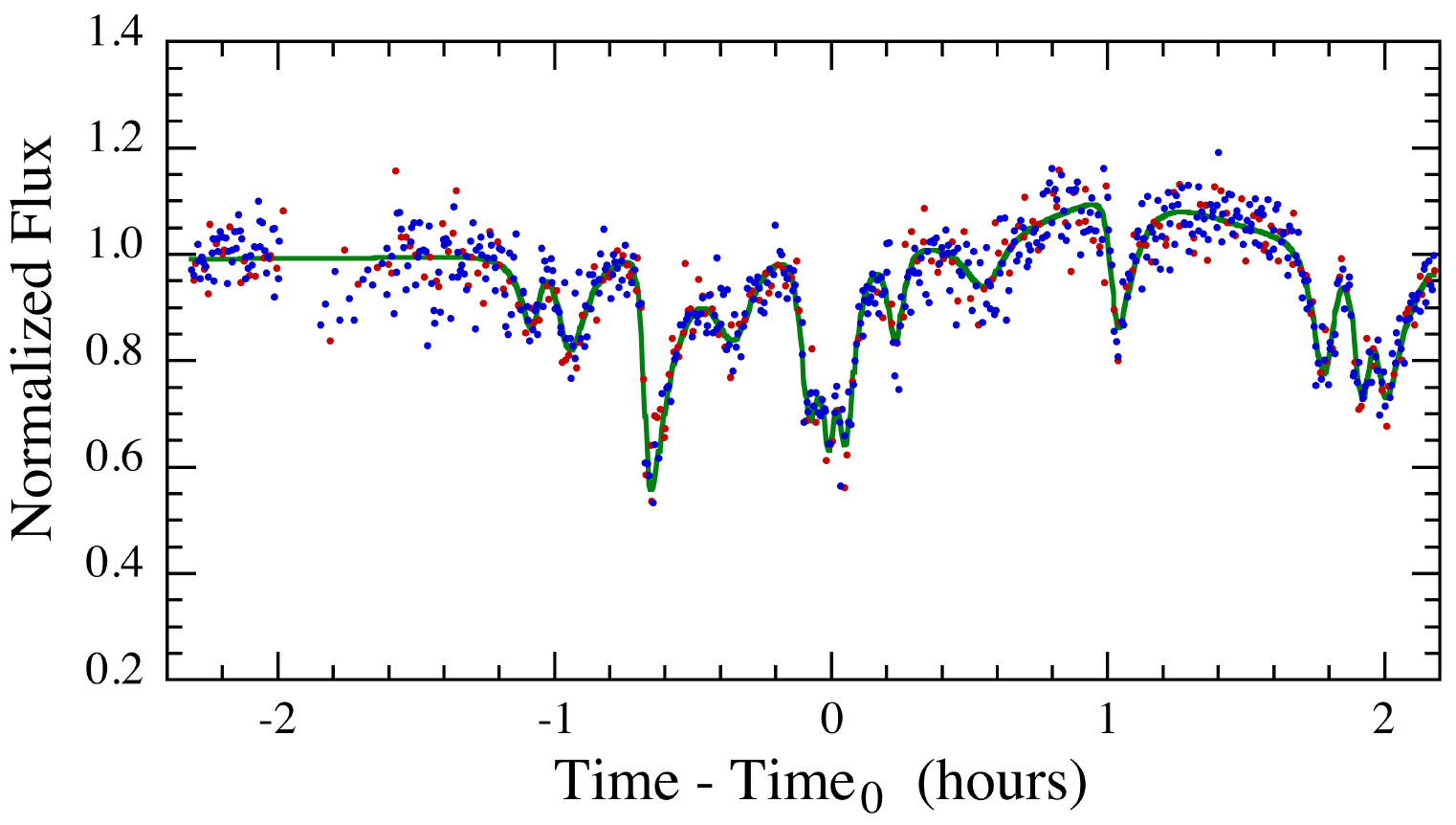}
\caption{Flux data from the night of 2015 Dec.~16th.  The red and blue points are from the Gary 14$''$ and Kaye 32$''$ telescopes, respectively.  Time$\,_0$ is the mid-phase as defined in Table \ref{tbl:periods} and propagated forward to this night.  The continuous green curve is the best fit of a function of the form of Eqn.~(\ref{eqn:ahs}).  We utilize these fits to determine the times and depths of the dips.}
\label{fig:fluxDec16}
\end{center}
\end{figure}

From an inspection of the raw flux data in Fig.~\ref{fig:data}, as well as from the folded light curve in Fig.~\ref{fig:fold1}, it is clear that there are numerous large dips occurring in this object at many different phases of the ``A'' period.  Therefore, in order to understand how the dip features evolve from night to night, we adopted the following approach which is illustrated in Fig.~\ref{fig:fluxDec16}.  Here we plot the flux data from the single night of 2015 Dec 16 obtained by two observers.   The flux data are then fitted to a non-physical model consisting of a set of asymmetric hyperbolic secants:
\begin{eqnarray}
{\rm Flux} = F_0 - \sum_m 2 \Delta F_m \times \left[{\rm exp}\left(\frac{t-t_m}{\tau_{\rm eg}}\right) + {\rm exp}\left(-\frac{t-t_m}{\tau_{\rm in}}\right)\right]^{-1}
\label{eqn:ahs}
\end{eqnarray}
where $m$ is the number of dips that are required to provide a good fit, and $\tau_{\rm eg}$ and $\tau_{\rm in}$ are the characteristic egress and ingress times for each of the dips (see Rappaport et al.~2014; Croll et al.~2015).  The fits were carried out with a Levenberg-Marquardt algorithm (Levenberg 1944; Marquardt 1963).  The times of the dipping events are taken from the values of $t_m$ and their depths from the $\Delta F_m$'s.  The fitting process was halted when all dips that appeared to be significant by eye had been found; we roughly estimate that this equates to all dips with 4-$\sigma$ significance or greater. 

\begin{figure*}
\begin{center}
\includegraphics[width=0.8 \textwidth]{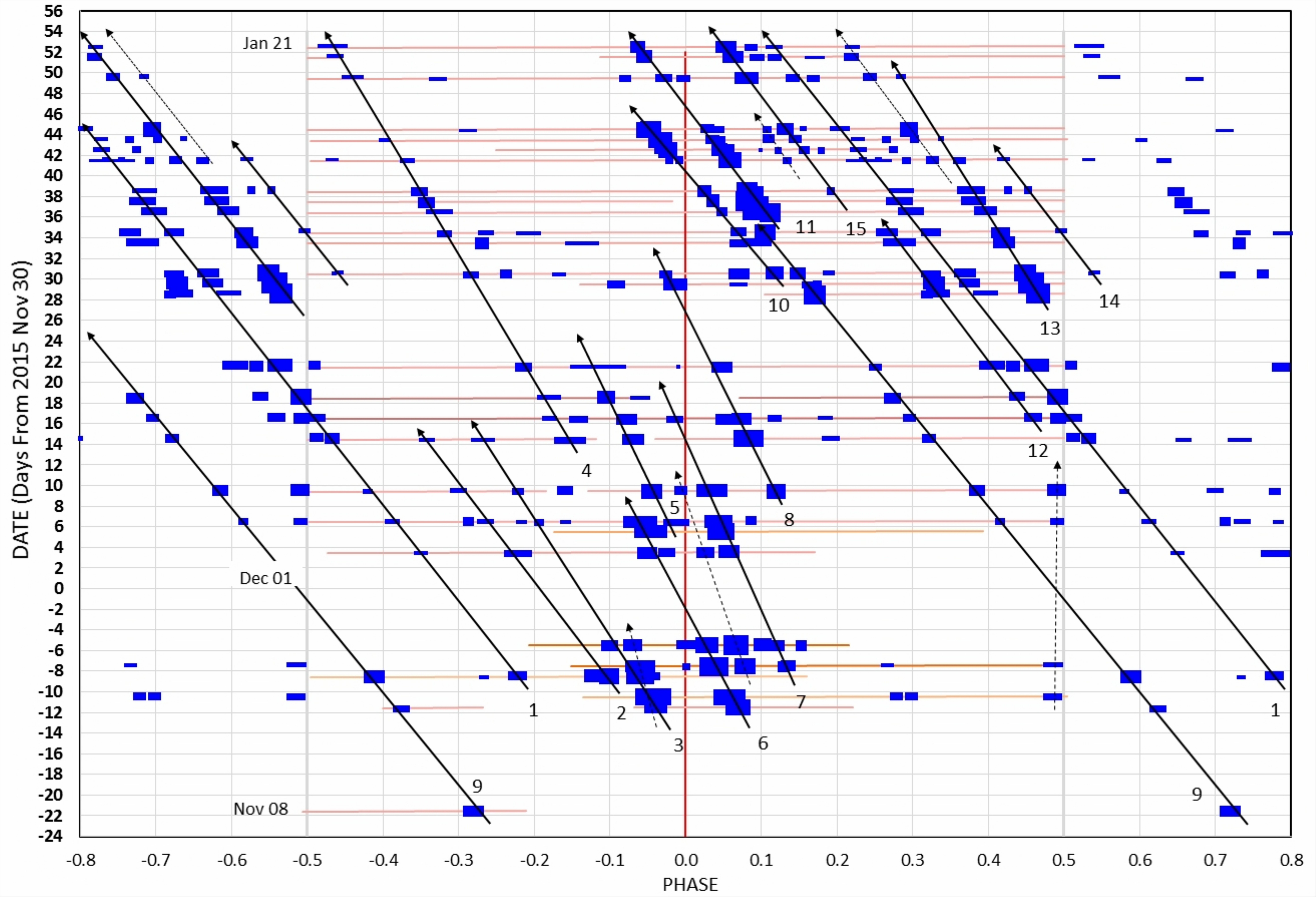}  \vglue0.3cm
\includegraphics[width=0.8 \textwidth]{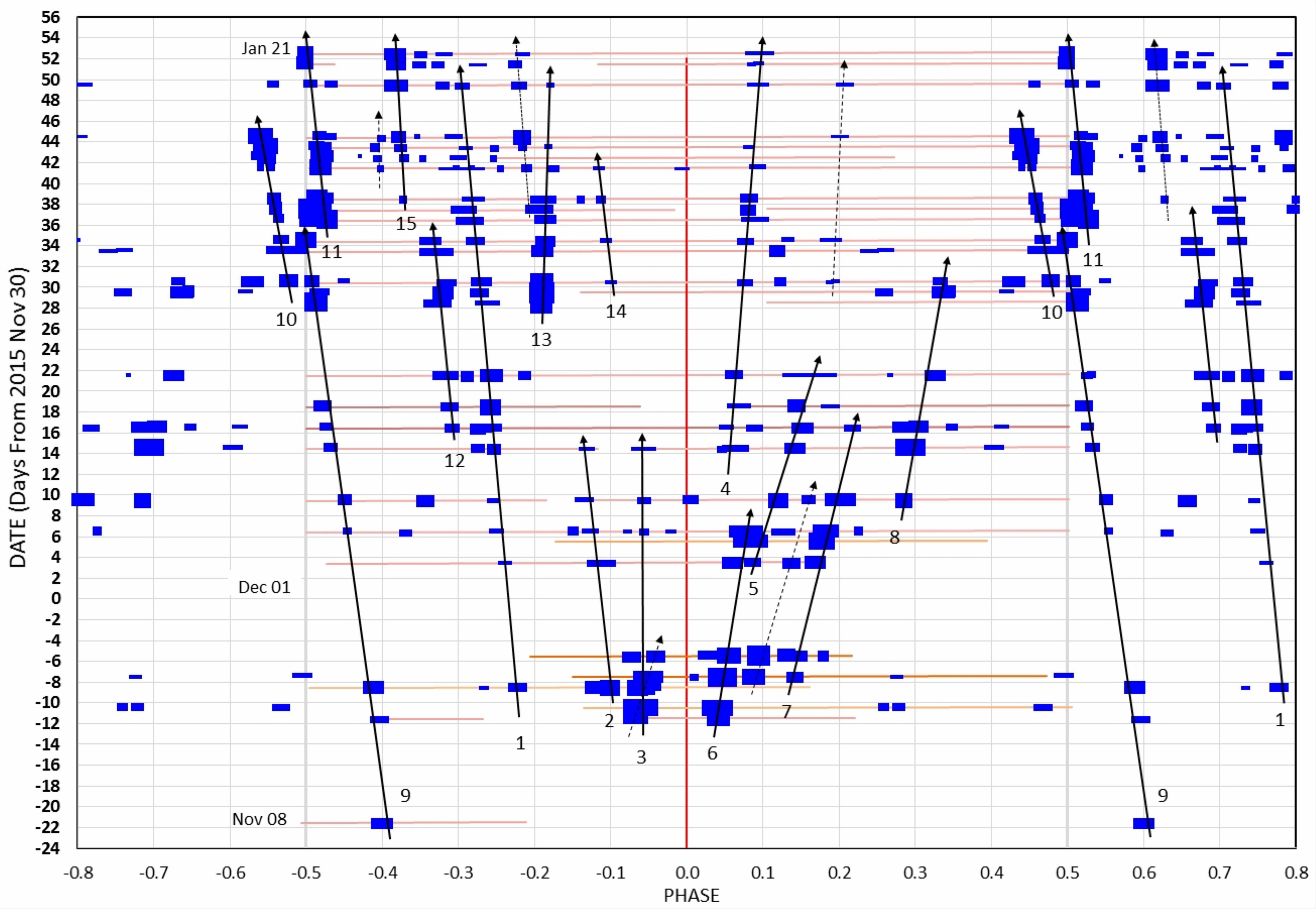}
\caption{Schematic, but quantitatively accurate, rendition of the flux time series on a nightly basis for the complete data set.  The data cover the interval 2015 Nov 8 through 2016 Jan 21.  We refer to this type of plot in the text as a ``waterfall diagram''.  The length and thickness of each blue segment linearly represents the duration (1--10 minutes) and depth (5-50\%) of the individual dips in flux.  The top panel shows the diagram when phased to the best-fit ``A'' period of 4.5004 hours.  The slopes of the solid lines connecting the same feature over different nights indicate the drift rates of the dust regions with respect to the ``A'' asteroid.  The bottom panel shows the same diagram, but phased to the mean drift period seen in the top panel.  Note the small, but significant, residual level drifting in different directions.  This demonstrates that the drift period is not a stable feature of the system. The faint dotted lines are used to indicate tentative patterns, one of which has a zero slope.  The faint red horizontal lines connect the dips on a given night. }
\label{fig:waterfall}
\end{center}
\end{figure*}

We then use the derived times and depths of the various dips to construct a ``waterfall'' diagram to show how the phases of the dips, with respect to the ``A'' period, are changing from night to night.  The top panel of Fig.~\ref{fig:waterfall} shows such a waterfall diagram constructed for the complete data set using the ``A'' period for the purpose of phasing the night-to-night dipping patterns.  The data from each night are represented by a schematic diagram where the duration of the dip is indicated by the length of the blue segment, while the thickness of the symbol is proportional to the dip depth.  We then connect the dips which we believe correspond to the same object, but observed on different nights.  The vast majority of the dips are clearly drifting with respect to the nominal fold period (``A'') by an amount that corresponds to $\sim$$2.5 \pm 0.2$ minutes per day.  The RMS dispersion in this value is 0.5 minutes per day for the 15 linear drift patterns which exhibit this approximate drift rate.  

In spite of the drifting nature of the features, we can see visually why the ``A'' period shows up in the BLS transform.  This is due to the fact that a substantial fraction of the dips, including most of the deeper ones, occur between phases $-0.1 \rightarrow +0.1$.  
	
Typically, a constant phase drift simply corresponds to an incorrect assumption about the period against which the data have been phased up.  Two and a half minutes per day corresponds to a ``adjusted'' period of 4.493 hours.  We also show in the bottom panel of Fig.~\ref{fig:waterfall} a second waterfall plot, but this one is phased to the mean drift period.  One can now see that there are clear sets of dipping events that become `stacked' vertically, or nearly so.  The fact that there are still some residual drifting motions using this period indicates that the mean drift period is not a precisely constant period of the system.  Furthermore, we see no evidence for that period in any of the search algorithms we employed.  There are two possible reasons for this: (1) there are so many dips over all orbital phases that the folded light curve does not develop enough of a depression at a specific orbital phase to reveal itself; and/or (2) this is not a long-term stable period, but in fact is fundamentally related to the ``A'' period itself.  There is clear evidence for the former hypothesis  in the bottom panel of Fig.~\ref{fig:waterfall}.  We elaborate on the latter possibility in the next section.  

\begin{deluxetable}{lc}
\centering
\tablecaption{Summary of Periods}
\tablewidth{0pt}
\tablehead{
\colhead{Descriptor} &
\colhead{Period or Epoch\tablenotemark{a}} 
}

\startdata
K2 ``A'' period\tablenotemark{b} & $4.4989 \pm 0.0001$ \\
Epoch of phase 0 (K2 ``A'')\tablenotemark{b} & $2456810.307 \pm 0.001$ \\
``A'' period -- this work &  $4.5004 \pm  0.0013$  \\
Epoch of phase 0 (this work) & $2457316.686 \pm 0.005$ \\
\hline
drift 1\tablenotemark{c}  &  4.4951$\pm 0.0006$ \\
drift 2\tablenotemark{c} &  4.4946$\pm 0.0005$ \\
drift 3\tablenotemark{c} &  4.4943$\pm 0.0008$ \\
drift 4\tablenotemark{c} &  4.4943$\pm 0.0003$ \\
drift 5\tablenotemark{c} &  4.4939$\pm 0.0008$ \\
drift 6\tablenotemark{c} &  4.4936$\pm 0.0006$ \\
drift 7\tablenotemark{c} &  4.4935$\pm 0.0003$ \\
drift 8\tablenotemark{c} &  4.4928$\pm 0.0005$ \\
drift 9\tablenotemark{c} &  4.4918$\pm 0.0005$ \\
drift 10\tablenotemark{c} &  4.4914$\pm 0.0005$ \\
drift 11\tablenotemark{c} &  4.4914$\pm 0.0006$ \\ 
drift 12\tablenotemark{c} &  4.4914$\pm 0.0003$ \\
drift 13\tablenotemark{c} &  4.4913$\pm 0.0004$ \\
drift 14\tablenotemark{c} &  4.4910$\pm 0.0002$ \\
drift 15\tablenotemark{c} &  4.4905$\pm 0.0008$ \\
inferred mean drift period\tablenotemark{d} &  $4.4928 \pm 0.0016$   \\
\hline
C15\tablenotemark{e} & $4.4912 \pm 0.0004$ \\
\hline
G15a\tablenotemark{f} & $4.49337 \pm 0.00021$ \\
G15b\tablenotemark{f} & $4.49252 \pm 0.00011$ \\
G15c\tablenotemark{f} & $4.49257 \pm 0.00052$ \\
G15d\tablenotemark{f} &  $4.49355 \pm 0.00040$ \\
G15e\tablenotemark{f} & $4.49110 \pm 0.00006$ \\
G15f\tablenotemark{f} & $4.49513 \pm 0.00046$ \\
G15 mean drift period\tablenotemark{f} & $4.4930 \pm 0.0013$ \\
\hline
K2 ``B''\tablenotemark{b}  &   $4.6053 \pm 0.0001$    \\
K2 ``C''\tablenotemark{b}  & $4.7828 \pm 0.0001$  \\
K2 ``D''\tablenotemark{b}  & $4.5500 \pm 0.0001$\\
K2 ``E''\tablenotemark{b}  & $4.8234 \pm 0.0001$ \\
K2 ``F''\tablenotemark{b}   & $4.8584 \pm 0.0001$ 
\enddata
\tablenotetext{a}{Periods are in hours and dates are in BJD.}
\tablenotetext{b}{Derived from the K2 observations; Vanderburg et al.~(2015).}
\tablenotetext{c}{The error bars are the accuracy with which we can derive the {\em offset} from an assumed well-defined period of 4.5004 hours.}
\tablenotetext{d}{Derived from the weighted mean and RMS standard deviation of the 15 drift periods listed above.}
\tablenotetext{e}{From Croll et al~(2015).}
\tablenotetext{f}{From G\"ansicke et al.~(2015).}
\label{tbl:periods}
\end{deluxetable}

In addition to the increased activity near zero phase in the top panel of the waterfall plot (Fig.~\ref{fig:waterfall}), it can be seen that each drifting object's transit activity, as measured by transit depth, varies with date. For example, the long-lived object \#9 is first observed with a depth of 33\%, and dwindles to $\sim$15\% for over a month, then becomes active at the end of 2015 December when its depth reaches $\sim$60\%, and finally it fades to disappearance a few days later. Each drifting object's transit depth has its own depth history.

\begin{figure}
\begin{center}
\includegraphics[width=1.01 \columnwidth]{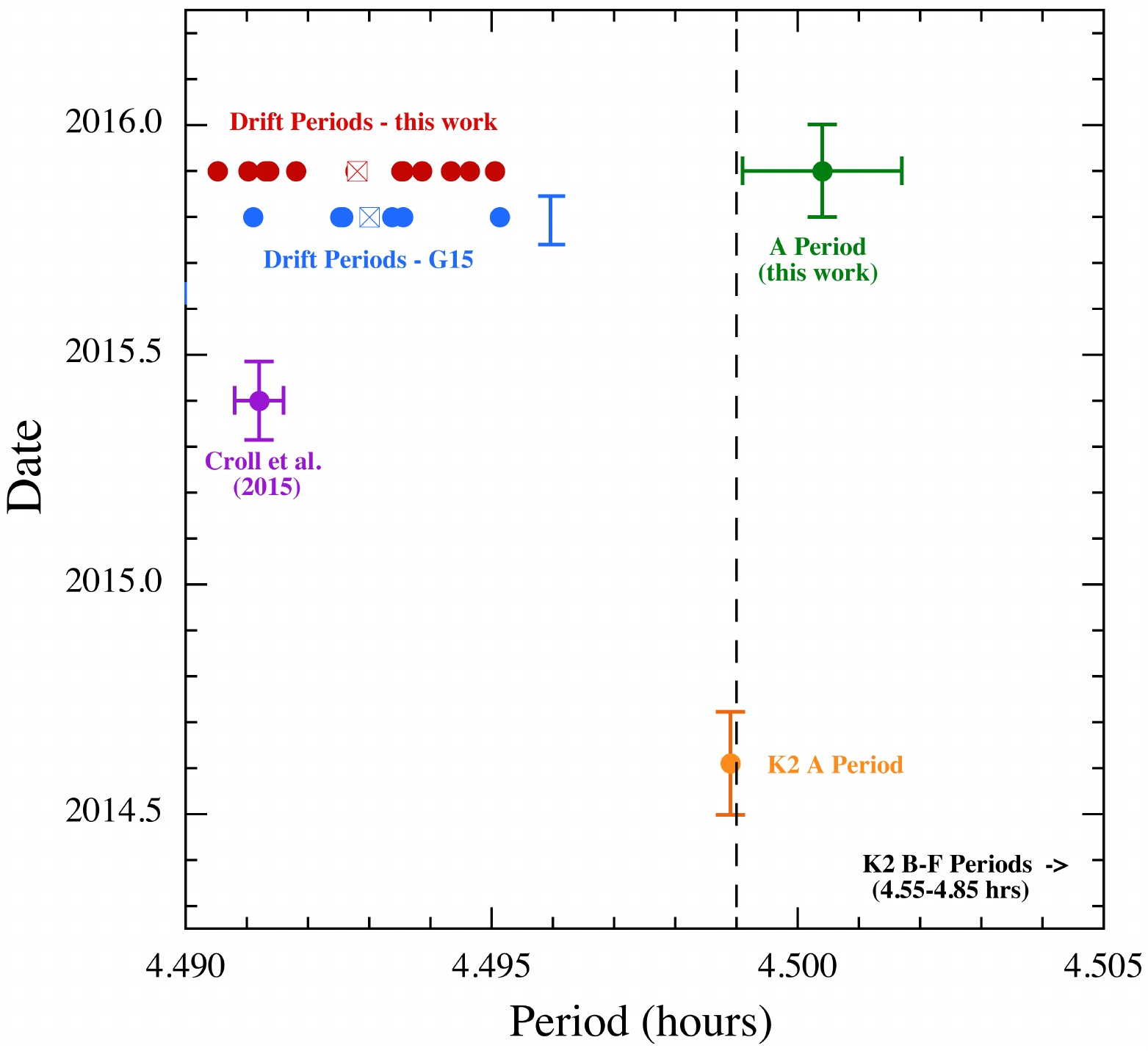}
\caption{A summary of the periods measured to date for WD 1145+017 that are near the ``A'' period at $\sim$4.5 hr. The ``A'' period itself as measured with K2 (orange circle) and in this recent work (green circle) are shown in mid 2014 and near 2016, respectively.  The vertical dashed line at 4.499 hours is consistent with both measurements to within their mutual uncertainties.  The K2 periods ``B'' through ``F'' are far off the plot to the right.  The drift periods measured by Croll et al.~(2015), G\"ansicke et al.~(2015), and in the current work are shown as purple, blue, and red circles respectively, at shorter periods to the left; a number of symbols representing closely spaced periods cannot be resolved.  The open square in each of the latter two drift period sets represents the mean of the drift periods. The vertical ``error bars'' represent the duration of the observations; the range for the red points is given by the green bar at right (i.e., the same for all the data in this paper).  The times of the G\"ansicke et al.~(2015) drift periods and those of the present work have been shifted down and up, respectively, by 10 days for clarity.}
\label{fig:periods}
\end{center}
\end{figure}

We summarize in Fig.~\ref{fig:periods} the periods measured to date for WD 1145+017 that are near the ``A'' period at $\sim$4.5 hr. We show both the ``A'' period itself as measured with K2 approximately one and a half years earlier as well as the value we determine in this work.  We draw a dashed vertical  line at 4.499 hours that is consistent with both measurements to within their mutual uncertainties.  Note that the K2 periods ``B'' through ``F'' are far off the plot.  We interpret the periods measured by Croll et al.~(2015) and G\"ansicke et al.~(2015) as drift periods under our model; the drift periods measured by Croll et al.~(2015), G\"ansicke et al.~(2015), and in the current work are shown as purple, blue, and red points respectively, all at distinctly shorter periods than the ``A'' period.  We also mark the mean values of the drift velocities with different symbols.

An inspection of Table \ref{tbl:periods} and Fig.~\ref{fig:periods} shows considerable overlap between our drift periods and the six periods that G\"ansicke et al.~(2015) list in their Table 1.

Finally, we note that we can use the linearity of the drift features (i.e., a track of phase, $\phi$, vs.~time) in Fig.~\ref{fig:waterfall} to set interesting constraints on their orbital decay lifetimes. For the longest lasting tracks (i.e., \#1 and \#9), the limit on the quadratic term in formal fits is $\ddot \phi \lesssim 6 \times 10^{-6}$ cycles day$^{-2}$.  The corresponding inferred dynamical lifetimes, $P_d/\dot P_d \equiv 1/(P_d \ddot \phi$), where $P_d$ is the drift period, are over 1000 years. Even for some of the shorter tracks seen in Fig.~\ref{fig:waterfall} the implied dynamical lifetimes are over 100 years.  Of course, if the orbiting object completely disintegrates in a much shorter time, then the longer dynamical lifetime has no meaning.  However, these dynamical lifetimes do suggest that the orbits are not significantly decaying while they are emitting dust.
	
\section{Interpretation of the Observations}
\label{sec:interpret}

From a perusal of our observational results, we reach a number of empirical conclusions.  First, there is only one period that is securely detected and  appears not to be transient, and that is $4.5004 \pm 0.0013$ hours, which is consistent with the ``A'' period detected during the K2 observations.
See Table \ref{tbl:periods} for a summary of all the periods associated with WD1145+017.  Second, $\sim$15 dip features drift with respect to the ``A'' period.  The orbital periods of these non-permanent features of the system are {\em all} shorter than the ``A'' period by an average of about 1/2 minute (out of an orbital period of 270 minutes), which agrees with the findings reported in Croll et al.~(2015) and G\"ansicke et al.~(2015).  As far as we can determine, the drift rates of the different features are rather similar to each other, but are not all exactly the same (see Fig.~\ref{fig:waterfall}) with a dispersion of $\sim$20\%.  Third, none of these shorter periods shows up robustly in any of the transforms; but, to be fair, they are not, strictly speaking, `resolvable' in a data set that is only $\sim$64 days long (the interval when observing sessions provided complete orbital phase coverage, or nearly so).  There is, however, sufficient accuracy in determining the periods that it would be straightforward to distinguish between the ``A'' period itself and the drift periods if {\em only one or the other} were present. Fourth, at least some of the drifting dip features in the light curve seem to have limited lifetimes, perhaps of the order of a few weeks\footnote{The alternative to this interpretation is that the transit depths of these fragments just become too small to be detected.}.  Finally, many of the individual dip features exhibit asymmetric profiles with the egress time generally longer than the ingress times, somewhat reminiscent of the transit patterns exhibited by the planets with dust tails: KIC 1255b and KOI 2700b (Rappaport et al.~2012; 2014; Brogi et al.~2012; Budaj 2013; see Sect.~\ref{sec:discuss} for details), and described in Vanderburg et al.~(2015) and Croll et al.~(2015).  

We propose a relatively simple strawman model that may explain many of the observed features of the WD 1145+017 lightcurve.  The basic idea is that there is a substantive asteroid (or planetesimal), hereafter ``A'', in an orbit that is stable over intervals of years, if not longer.  The mass of this asteroid, $M_a$, must lie in the range $5 \times 10^{22} \lesssim M_p \lesssim 10^{24}$ grams, based on a number of arguments that we present below (see also Vanderburg et al.~2015).  We further propose that the asteroid is virtually filling its critical potential surface, and discrete chunks of matter (hereafter referred to as `fragments') episodically break off of the asteroid at the L1 point and go into a slightly smaller orbit.  We compute analytically the expected periods of these smaller bodies, and the concomitant drift periods with respect to the ``A'' period, in Appendix \ref{app:drift}.  We find that
\begin{eqnarray}
\frac{P_a-P_f}{P_a} \simeq 2 \times 3^{2/3} \mu^{1/3}
\end{eqnarray}
where $\mu$ is the ratio of asteroid mass to the mass of the white dwarf host star, and $P_a$ and $P_f$ are the orbital periods of the asteroid and the fragment, respectively.  

The L1 point is the natural location from which loose or weakly bound material can break off and fall away from the asteroid.  This, and the L2 point, are the only local saddle points around the critical potential surface, and once an object gets beyond L1 it is free to fall into an independent orbit about the star.

As the smaller released bodies move about the white dwarf they are heated further until there is a freely escaping flow of vapors comprised of the mineral content of the fragment.  This vapor would then condense into dust clouds after escaping the extremely low-gravity environment of the fragment.  Radiation pressure from the white dwarf may then push some of the dust into a trailing tail configuration reminiscent of solar system comets (see Rappaport et al.~2014 and Vanderburg et al.~2015 for details).  Such dust tails would explain the longer egress times compared to the ingress times; this occurs $\approx 70\%$ of the time, compared with $\approx 30\%$ of the time with longer ingress times than egress times (for details see Sect.~\ref{sec:discuss}).  

We adopt a rough mass-loss rate estimate of $\sim$$10^{11}$ g s$^{-1}$, which is $\sim$10 times higher than the estimate provided by Vanderburg et al.~(2015) due to the higher levels of obscuration at the current epoch than during K2.  With admittedly incomplete information, we can also estimate that these small released fragments last for only a few weeks before completely disintegrating (see Fig.~\ref{fig:waterfall}).  In that case, it follows simply from the product of the adopted value of mass-loss rate and an assumed $\approx$ 20-day lifetime that their masses should be of order of $2 \times 10^{17}$ g, or about the mass of Haley's comet (and a diameter of $\sim$6 km).  Finally, in this regard, we note that with broken off chunks of material this large, the asteroid itself still has sufficient mass to release about $10^6$ of these fragments.  We see about 6-10 of these objects in the system at any one time, and if they last for three weeks, that implies an average release rate of about one fragment every few days.  Thus, the expected lifetime of the asteroid would be $\sim$5,000 years (at the current activity level), or as much as 10 times longer if the K2 level of activity is typical.

We point out that no part of the above scenario precludes the existence of the five other asteroids (or planetesimals) which were inferred from the K2 data (``B'' through ``F'').  Those have periods that range from 1\% to 10\% {\em longer} than the ``A'' period.  These differences from the ``A'' period (on the longer side) are considerably greater than the differences between the ``A'' period and those we infer for the drifting fragments (on the shorter side).

Finally, we note that if bodies which cause periodic dips in flux with period $P_f$ (i.e., the fragments) are launched into their orbits at phases synchronized with the transits of the body which launched them (i.e., the asteroid, with period $P_a$), then it is an interesting question as to which period ultimately shows up in a periodogram made from the resultant flux data.  We reason that if the fragments last for a duration that is short compared to that of the observations, and there are typically a couple fragments in the system at any one time, then it is $P_a$ which predominantly appears in periodograms.  Conversely, if the fragments last for a time comparable to the observations, but there are still only a few in the system at any given time, then it will be $P_f$, the period of the fragments, that will show up.  If there are too many fragments in the system at any one time, then both signals will tend to be washed out.  This is an interesting mathematical problem to be worked out analytically, but one that is beyond the scope of the present paper.

\begin{figure}
\begin{center}
\includegraphics[width=0.99 \columnwidth]{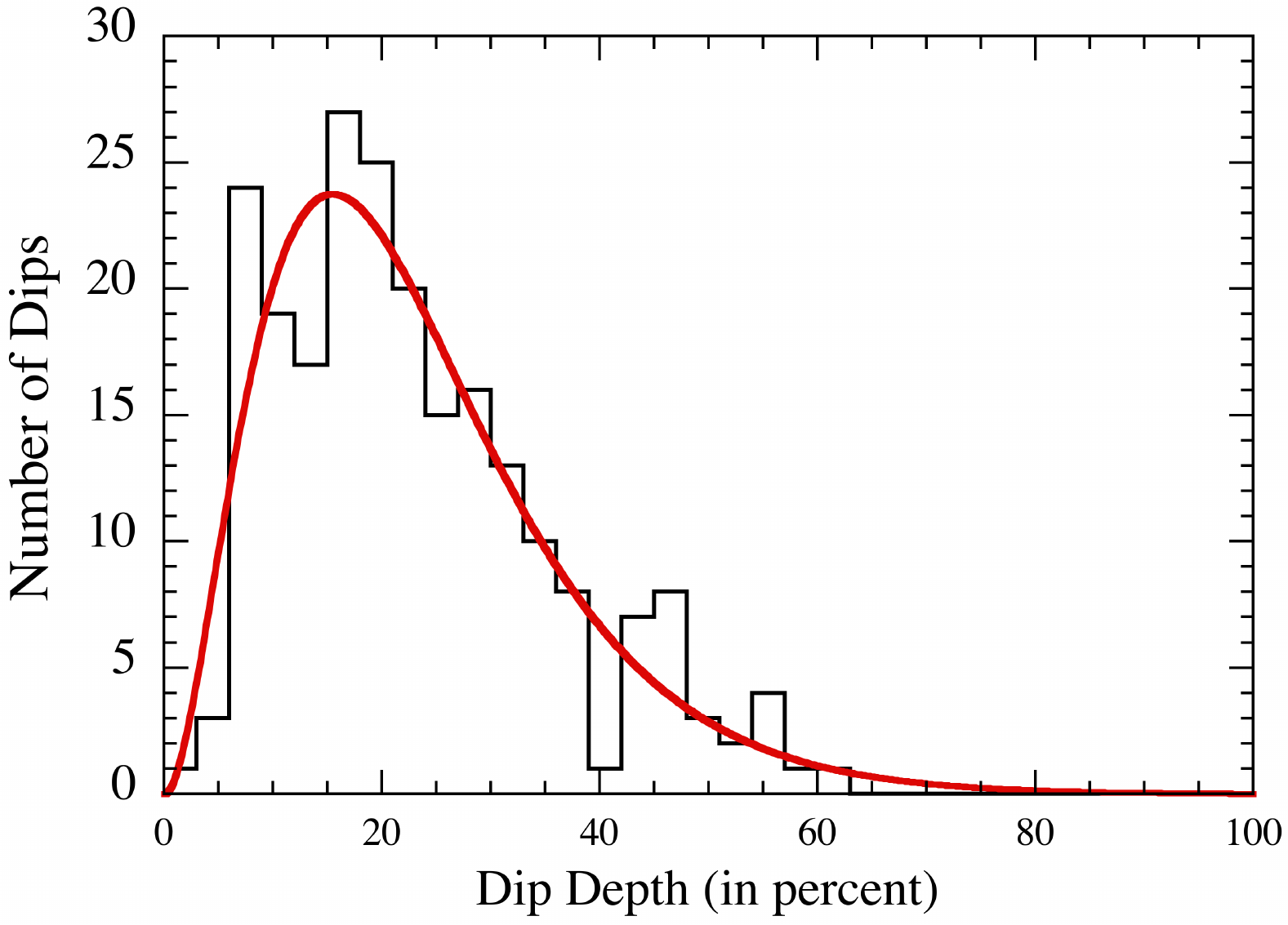}
\caption{Distribution of the depths of 227 dips measured during the course of our 2.5-month long observing campaign of WD 1145+017.  The red curve is a fit to an ad hoc analytic expression of the form: $\mathcal{P}(D) \propto D^2 {\rm exp}(-D/D_0)$ where $D$ is the dip depth and $D_0$ is a fitted parameter found to be 7.73\%.  The sharp curoff below a depth of $\sim$6\% is very likely imposed by the limitations of our telescope apertures. }
\label{fig:depths}
\end{center}
\end{figure}

\section{Discussion}
\label{sec:discuss}

In this work we have identified some 237 significant dips in the intensity of WD 1145+017 that are presumably due to obscuration by dust that is local to   bodies orbiting around the white dwarf.  The distribution of dip depths is shown in Fig.~\ref{fig:depths} along with a fitting function.  The falloff below 15\% in the fitting function, and a shaper cutoff below depths of $\sim$6\% are undoubtedly due to the limitations of our relatively small aperture telescopes and the lack of S/N for detecting shallow dips. This was especially true under the unfavorable observing conditions that characterized the 2015 November and early 2015 December observing sessions. Thus, whereas the number of dips we typically found per orbit was 6 to 10, the actual number that might be observed with a larger aperture telescope is potentially substantially greater. This remains as a task for future observers (see also G\"ansicke et al.~2015).

In spite of the limitations of our telescope apertures, we were able to track about 15 apparently independent objects orbiting the white dwarf with periods near 4.493 hours (see also G\"ansicke et al.~2015).  These are best visualized in Fig.~\ref{fig:waterfall}.  The periods of these objects are in the range of 4.490--4.495 hours, which are all shorter than the ``A'' period of 4.500 hours by only $\sim$0.2\%.  Nonetheless, all these periods are sufficiently close that we were led to the conclusion that they are all related.  Our tentative hypothesis for explaining the relation among these periods is that there is a substantive, long-lived body in the 4.500-hour orbit (the asteroid), which loses fragments of matter that are a tiny fraction of its total mass, and that naturally go into slightly shorter period orbits (see Appendix \ref{app:drift}).  These smaller fragments last for only a matter of days or weeks before they are destroyed via evaporation and sublimation induced by heating from the central star.  

We reasoned that if these fragments first appear as dust emitting objects in phase with the parent asteroid, and last for only a short time, during which they do not drift far in orbital phase from the asteroid, then the period which shows up in periodograms will be the K2 ``A'' period.  If, by contrast, these fragments last for a substantial time (e.g., months), and there are many of them, then it is likely that only a weak signal, or no signal, will appear in the periodogram at the ``A'' period. In addition, if there are too many drifting objects then it will be difficult to detect the drift period either because these objects tend to populate all orbital phases and, moreover, there appears to be a dispersion in drift rates.  In this work we found a convincing signal at the ``A'' period. That fact, in addition to a number of the above listed circumstances, can be partially understood by examination of Fig.~\ref{fig:waterfall}.  In the top panel, phased to the ``A'' period, the preponderance of deep dips do occur between phases $-0.1 \rightarrow +0.1$.  By contrast, in the bottom panel, phased to the mean of the drift periods, the dips appear too uniformly over all phases to allow that period to show up in periodograms.  

The behavior of the drifting fragments seen in Fig.~\ref{fig:waterfall} raises a number of generally interesting questions in the context of our proposed scenario for the dips: (i) Do fragments turn on (i.e., start to emit enough dust to produce noticeable dips) immediately after breaking off from the L1 point of the asteroid?  (ii) If not, how long can a fragment remain inactive until it turns on?  (iii) Once a fragment turns on, how long can it remain active?  Of course each fragment will be different, but since we have 15 cases for study it has been possible to place bounds on the answers.  In the following we give some illustrative answers.

With respect to the first two questions, if a fragment `turned on' (i.e., started emitting dust) immediately upon breaking off from the L1 end of the asteroid, it would produce dips starting at the asteroid's phase location. For example, tracks \#7, \#8, \#10, and \#11 appear to start at phase $\approx 0.13$ (see top panel in Fig.~\ref{fig:waterfall}), and persist for about three weeks each.  During this time, the phase drifts to the left in Fig.~\ref{fig:waterfall} (top panel) by $\approx 0.17$.  If these represent fragments that started emitting dust immediately after breaking away from the asteroid, then we would interpret this as pointing to the asteroid being located at phase 0.13.  In this case tracks \#2, \#3, and \#5, which first appear only at more negative phases, would be inferred to have `turned on' only some three weeks after their release, while fragment \#6 would have been delayed by only 6 days before turning on.  Fragment \#13 begins at phase +0.47, and it has a drift rate of $-$2.18 min/d, so projecting this line all the way back to a phase of +0.13 would require $\sim$82 days. In other words, this fragment could have remained inactive for over four months from its hypothesized 2015 Oct 05 launch from L1.

In regard to the third question listed above, we note that the longevity of activity can also be estimated from the cases of the 15 drifting fragments. The median longevity is at least 20 days, and maximum longevity is at least 58 days (fragments \#1 and \#9). Even these durations may be underestimates, since it appears that a fragment can become too inactive to produce a dip that is detectable using the modest telescopes of this study, and later return to activity (e.g., fragments \#5, 6, 7 and 12). This matter remains to be studied by future observations.

Finally, examination of the bottom panel in Fig.~\ref{fig:waterfall} gives the impression that a number of the drift lines may be diverging from one or more common points several months in the past.  If those projections converge at common times near phase zero (in the ``A'' frame) then we could consider this to be evidence that one or more groups of fragments were created on those dates. At the present time, we do not have the temporal coverage or precision to carry out such projections.  Future observations, however, can be viewed with such a model in mind, and may allow one to determine the date of creation of one or more groups of fragments by looking for these points of convergence of the drift lines near phase zero of the main body. 

\begin{figure}
\begin{center}
\includegraphics[width=0.99 \columnwidth]{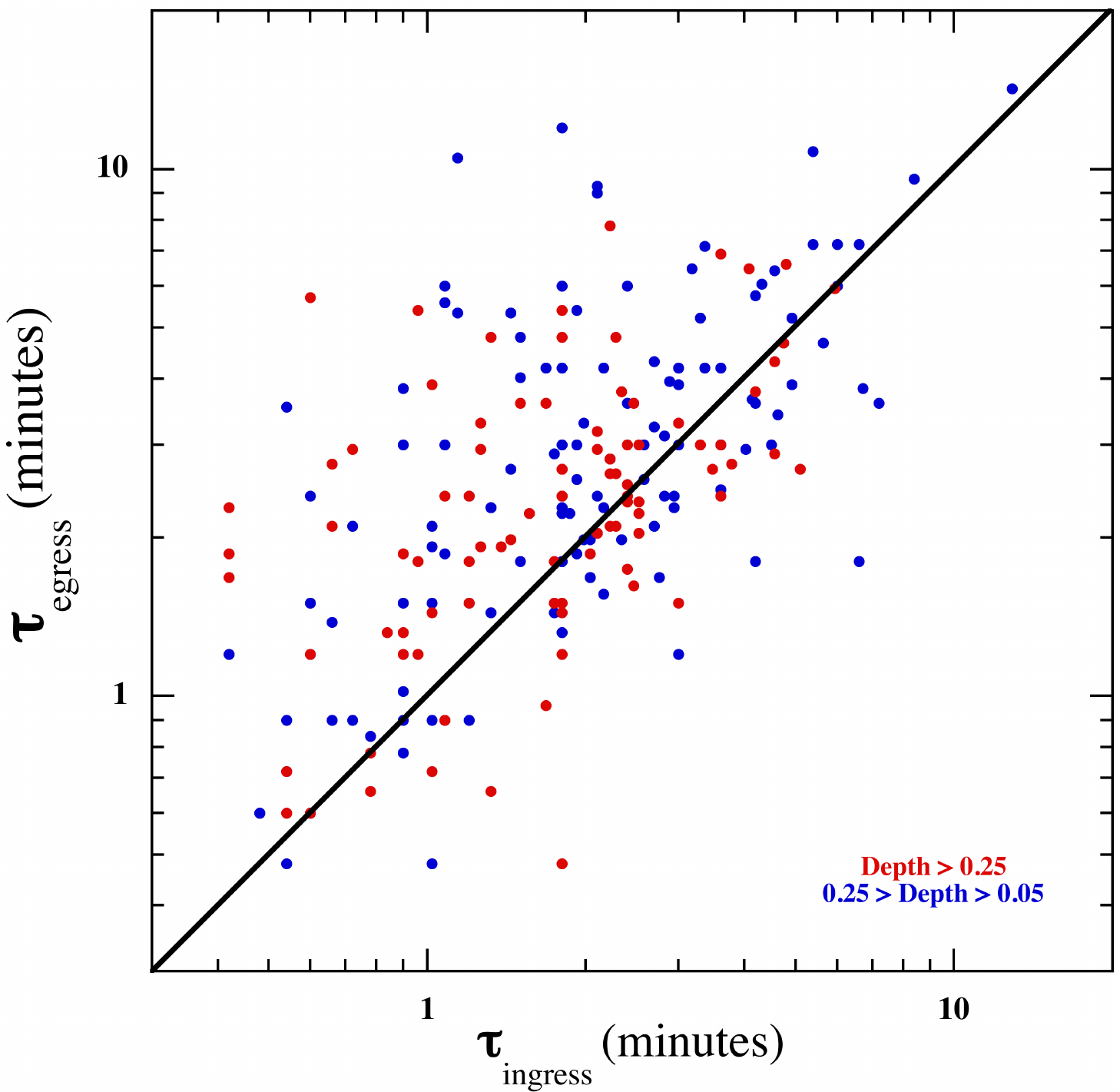}
\caption{Correlation between the ingress and egress times, $\tau_{\rm in}$ and $\tau_{\rm eg}$ (see Eqn.~\ref{eqn:ahs}), respectively, of all dips with a depth $\gtrsim 5\%$.  There are 140 dips with $\tau_{\rm eg} > \tau_{\rm in}$; 61 with the reverse inequality, and 23 with $\tau_{\rm eg} \simeq \tau_{\rm in}$ (to within 5\%). }
\label{fig:tau1tau2}
\end{center}
\end{figure}

The drifting objects (the `fragments') have a notable tendency to show longer egress tails than ingress tails.  In Fig.~\ref{fig:tau1tau2} we show a correlation plot between the fitted time of egress vs.~that for ingress.  After eliminating 12 dips with depths of $\lesssim 5\%$, we find 140 dips with $\tau_{\rm eg} > \tau_{\rm in}$ and 61 with $\tau_{\rm in} > \tau_{\rm eg}$ (a 5.5-$\sigma$ asymmetry).  The longer egress tails, in this picture, would naturally correspond to dust tails that trail the fragment due to radiation pressure (see, e.g., Rappaport et al.~2014).  However, as was shown in Sanchis-Ojeda et al.~(2015), dust tails can also {\em lead} the emitting object if the radiation pressure is low enough or if the dust particles have a small scattering cross section (see, e.g., Kimura et al.~2002).  Finally, in regard to the results in Fig.~\ref{fig:tau1tau2}, we note that we have not looked carefully into possible correlated errors between $\tau_{\rm eg}$ and $\tau_{\rm in}$, and that further analyses may be required to confirm these apparently significant and suggestive results.

We note that it takes very close to 1 minute for a point-like object to transit across the disk of the white dwarf (at orbital speeds of 315 km s$^{-1}$).  Therefore, the ensemble of ingress and egress times implies dust regions that range from the same size as the white dwarf, or smaller, all the way up to dust clouds that are an order of magnitude longer than the white dwarf's diameter.  

An important finding of this work, and a prediction of our model, is that the drifting fragments should {\em all} have a period that is shorter than that of the parent body (the asteroid).  The reason follows from the fact that in this scenario the fragments break off from, and leave, the L1 point (see Appendix \ref{app:drift}).  The question might be raised as to why the fragments always leave from L1 rather than L2, the two of which are both saddle points in the effective potential at essentially the same potential.  We suggest that the reason is the extreme heating of the tidally locked body (see Vanderburg et al.~2015) at the L1 point as opposed to the much lower temperatures that are likely on the dark L2 side of the body.  The higher equilibrium temperatures at L1 would tend to keep the surface material there either molten or at least more `plastic'.

Of the unexplained observational results for this object enumerated in the Introduction, the numerous transits observed in this work, coupled with the model we have proposed, go a long way toward providing satisfactory answers.  The significantly longer inferred duration of the transits (i.e., $\sim$50 min)
detected by Vanderburg et al.~(2015) in K2 data compared with the ground-based data of Croll et al.~(2015) and G\"ansicke et al.~(2015) can now be understood from the fact that the half-hour integration time of K2 is inadequate to track the numerous narrow dips and is mostly responsive to collections of narrow dips that congregate at a particular orbital phase.  The small differences in orbital period between the K2 ``A'' period and the drifting features found from ground-based observations are a consequence of the fact that fragments would naturally tend to fall into shorter-period orbits. The non-detection of the five weaker periods found by Vanderburg et al.~(2015) with K2 (``B''-``F'') during the Croll et al.~(2015) observations, as well as in this work, is likely explained by the fact that if those K2 signals continued at the same levels, they would be too weak to be easily detected with ground-based telescopes.  Finally, our scenario provides a natural explanation for how multiple occulting bodies can coexist in orbits that are very close to 4.5 hours.  In particular, they are hypothesized to originate from a single parent body, and their masses are too low to cause serious dynamical interactions.

One of the difficulties with the proposed model may be the high inferred mass loss rates from such small bodies.  Vanderburg et al.~(2015) showed that mass loss rates from a Ceres-size body ($\sim$1000 km in diameter), heated to an equilibrium temperature of $1500 \pm 100$ K, could be $\sim$$10^{10}$ g s$^{-1}$ in the form of vapors of some common minerals.  The fragments from this body, which we take to be of the order of $\lesssim 10^{-6}$ times the mass of Ceres, would then be two orders of magnitudes smaller in linear dimensions.  That would indicate that the fragments would have only $\sim$$10^{-4}$ times the surface area of Ceres, and therefore we might expect orders of magnitude lower values of mass-loss rate.  One possibility is that the fragments contain a portion of volatiles which act in a more explosive manner when embedded in a low-mass fragment that is suddenly released.  Of possible relevance to this is the fact that Xu et al.~(2016) have found O absorption lines in the circumstellar material near WD 1145+017, that may account for as much as 60\% of the mass fraction of the debris.  In addition to the presence of oxygen lines in the circumstellar material, there is also evidence for hydrogen in the otherwise helium dominated envelope (Xu et al.~2016).  While excess O has been interpreted as evidence for water (e.g., Farihi et al.~2013), Xu et al. (2016) caution that their O abundance is subject to a large uncertainty, and the H abundance is not unusually high for the $T_{\rm eff}$ of this white dwarf (Koester \& Kepler 2015).  Future observations could improve the precision of the O abundance on WD 1145+017 because Xu et al.~(2016) only detected oxygen lines in moderate resolution spectra.

A positive feature of our model for the drifting features is that the drift periods can be used to infer the mass of the parent asteroid with essentially no free parameters.  The down side of this is that there is no ready explanation for the spread in drift periods by $\sim$20\% with respect to the ``A'' period.  

We should also mention at least one other scenario which might explain some of the same observational facts.  There may be {\em no} `master body' in the system from which the dust emitting bodies break off and start an independent existence.  Rather, the `drifting objects' may be just a collection of debris in a narrow range of orbital periods with $\Delta P/P \simeq  0.2\%$ whose individual dust emission is erratic, and they can persist in those orbits for years.  Here we briefly examine this hypothesis.  In terms of orbital stability, a range in $\Delta P/P \simeq  0.2\%$ corresponds to a range in orbital radii of $\Delta a/a \simeq  0.13\%$.  If we require orbital separations of at least, say, 10 Hill's sphere radii for minimal stability over an interval of many years (especially given that there may be a dozen bodies), then we find that the mass of the fragments, $M_f$, must be:
\begin{equation}
M_f \lesssim 3\left(0.0013/10\right)^3 \,M_{\rm wd} \simeq 8 \times 10^{21} \, {\rm grams}
\end{equation}
for the case of circular orbits (eccentric orbits would substantially lower this limiting mass.)  This amounts to a body that is only 1\% the mass of Ceres.  If these bodies lose mass at an average rate of, say, $10^{10}$ g s$^{-1}$, then they could withstand complete disintegration for $2 \times 10^4$ years.  In this alternative interpretation, there are at least a dozen bodies in long-term stable orbits with periods covering a narrow range of $\sim$0.2\%  whose appearance simply depends on whether they happen to be emitting dust or not.  One argument against this scenario is that during two relatively lengthy  observations (80 days with K2 and the currently reported observations which span only somewhat less time), the only period that shows up in a periodogram to within 1\% of the ``A'' period is the ``A'' period itself, and not any of the slightly shorter periods which we identify as drifts.  If all the bodies had an equal status, as independent sources of dust, then this would seem highly unlikely.  

\section{Summary and Conclusions}
\label{sec:summary}

In this study we have observed WD 1145+017 on 37 nights over a two and a half month interval.  The observations were carried out at a combination of four privately operated amateur observatories equipped with telescopes of aperture 28 cm to 80 cm.  Some 237 significant dips in flux were observed during 192 hours of exposure.

The only period that shows up with significance in a BLS transform of the data is 4.5004 hours, which is within 3 parts in $10^4$ of the ``A'' period found by Vanderburg et al.~(2015) during the K2 observations.  We find some 15 dip features that repeat from one night to the next, and typically last for a few weeks.  The phases of these features drift systematically with respect to this ``A'' period at a rate of $\sim$2.5 minutes per day with a dispersion of $\sim$0.5 minutes per day.  The periods of the drifting features range from 4.4905 to 4.4951 hours and their mean period is 4.4928 hours.  

We propose that the drifting features are caused by dust-emitting fragments that have broken free of a more massive asteroid that orbits with the ``A'' period.  The underlying motivation for suggesting this model is that the most significant period detected during the K2 observations over 80 days was 4.4989 hours which is, to within the mutual uncertainties, a match for the only significant periodicity detected in periodograms of the current observations spanning only a somewhat shorter interval, but 1.5 years later.  It seems a highly likely conclusion that there is a relatively stably orbiting body in the system with this period.  The other drifting periods are the same, to within 0.2\%, of the ``A'' period, but seem not to be permanent features of the system.  We therefore are led to believe that the stable feature and the drifting features are physically related.

We show in detail how the orbital period of objects released from the L1 point of an asteroid go into slightly shorter orbital periods by a fractional amount that is proportional to the dimensionless Hill's sphere radius.  In turn, that radius is proportional to the mass ratio of the parent asteroid to the white dwarf raised to the 1/3 power.  We use this theoretical relation to derive an estimated mass for the parent asteroid of $\sim$$10^{23}$ g which is about 10\% the mass of Ceres.  
 
Through this work we have shown that small-to-modest sized telescopes can be extremely effective in following the evolution of the dips in WD 1145+017,  both in phase and depth.  While we have learned enough over the course of these 2.5-month-long observations to propose a tentative tidal fragmentation model for the complex array of dips, more monitoring of this object is needed to help confirm or refute the proposed scenario for the drifting features.  Moreover, we now know that the overall activity level of this source changes on timescales of months.  It could therefore be quite important to understand the longer timescale variations.

Our present observations show considerably increased activity compared to the K2 observations (Vanderburg et al. 2015) and the ground-based observations in early 2015 (Vanderburg et al. 2015; Croll et al. 2015). This nearly 10-fold increase in activity likely occurred sometime between the 2015 March-May observations (when the activity was still not at the current level; Vanderburg et al. 2015; Croll et al. 2015) and 2015 November (eight months later; see also G\"ansicke et al.~2015).  Something must have happened to cause this increased activity. Perhaps the ``A''-period asteroid had a collision with another object, such as a large fragment in an orbit whose eccentricity brought them together.

Importantly, in spite of the fact that we have assumed that dust is responsible for the dips in this source, there is no {\em direct} evidence currently available to demonstrate either the existence of azimuthally structured dust that can attenuate up to 60\% of the system light, or the properties of such dust (see, e.g., Croll et al.~2015). Therefore, we urge observations with larger telescopes covering as wide a wavelength range as possible to determine the wavelength dependence of the dips.

\acknowledgements 

We thank an anonymous referee for some very helpful and insightful comments. AV is supported by the National Science Foundation Graduate Research Fellowship, Grant No.~DGE 1144152.  TK thanks Cheryl Healy and Beth Christie for material support in telescope operations and equipment.  

\appendix

\section{A. Orbits of Asteroids Breaking Off Tidally Disrupted Planetesimals}
\label{app:drift}

In this Appendix we derive the orbits of fragmented bodies that break away from the L1 point of a larger asteroid orbiting a host star near its tidal breakup limit.  In particular, the goal is to establish the rate at which the broken-off bodies will gain in orbital phase with respect to the parent asteroid as a function of time.  In this derivation we utilize Roche's  (Hill's) geometry even though when the fragment is still near the asteroid the formalism is not exact (due to the extended and non-spherical nature of the body).  However, once the fragment goes into an independent orbit, the approximations should be excellent.  

\begin{figure}[h!]
\begin{center}
\includegraphics[width=0.5 \textwidth]{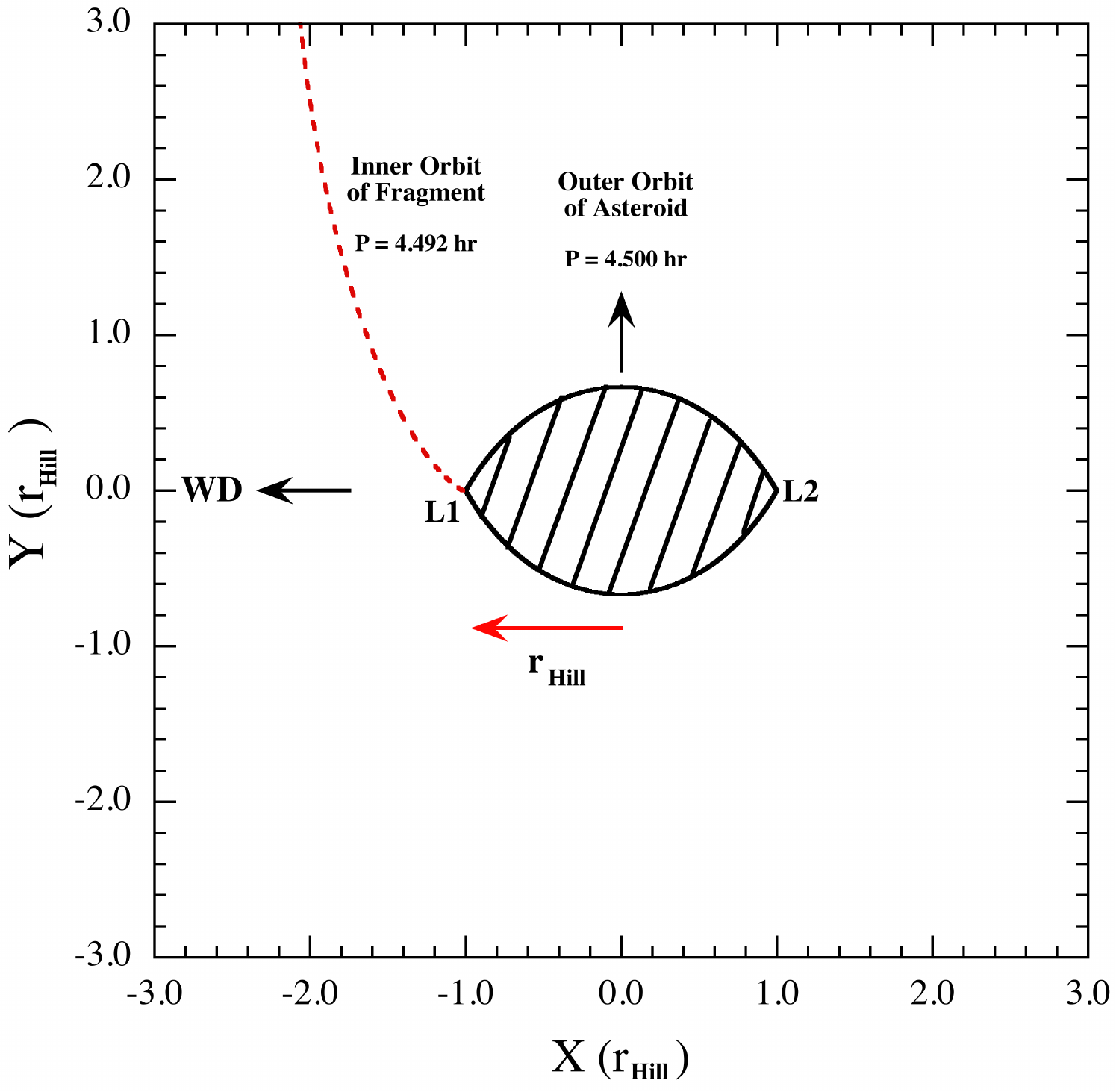}
\caption{Critical Roche surface in the $x-y$ plane, centered on an asteroid of low mass, $M_a$.  The L1 and L2 points are marked, as is the Hill's `sphere' radius.  The massive primary star (of mass $M_*$) is far to the left along the $x$ axis.  $X$ and $Y$ are in units of the dimensionless Hill sphere radius, which is given as $r_h = (\mu/3)^{1/3}$, where $\mu$ is defined as $M_a/(M_a+M_*)$.  The dashed red curve indicates the orbit of a fragment that breaks off from the L1 point and goes into an independent orbit.  The orbital period of the fragment will then be slightly shorter than that of the parent asteroid by a fractional amount that is proportional to $r_{\rm h}$}
\label{fig:Roche}
\end{center}
\end{figure}

The Roche potential expressed in cartesian coordinates about the center of mass of the orbiting asteroid is given by:
\begin{eqnarray}
\Psi(x,y) = -\frac{GM}{a} \left[\frac{(1-\mu)}{\sqrt{(x+1)^2+y^2}}+\frac{\mu}{\sqrt{x^2+y^2}}+\frac{1}{2} (1-\mu+x)^2+ \frac{1}{2} y^2\right] 
\end{eqnarray}
where $M_*$ and $M_a$ are the masses of the host star and asteroid respectively, $M = M_*+M_a$ is the total mass, $\mu = M_a/M$, $x$ and $y$ are the coordinates in the orbital plane, centered on the asteroid, in units of the orbital separation, $a$, and $G$ is Newton's gravitational constant.  The $x$ axis lies along the line joining the two bodies.  We will retain all terms of order $\mu^{1/3}$ and $\mu^{2/3}$, but drop terms of order $\mu$ (which are $\sim$$10^6$ and $10^3$ times smaller, respectively, than the $\mu^{1/3}$ and $\mu^{2/3}$ terms in this problem). The term with $\mu/(x^2+y^2)^{1/2}$ cannot be dropped because, even though $\mu$ is of third order, it is divided by terms of the order of $x$ and $y$ which go as $\mu^{1/3}$, and therefore the ratio goes like $\mu^{2/3}$. This results in:
\begin{eqnarray}
\Psi(x,y) = -\frac{GM}{a} \left[\frac{1}{\sqrt{(x+1)^2+y^2}}+\frac{\mu}{\sqrt{x^2+y^2}}+\frac{1}{2} (1+x)^2+ \frac{1}{2} y^2\right]
\end{eqnarray}
We now expand this expression in a Taylor series to second order in $x$ and $y$:
\begin{eqnarray}
\Psi(x,y) = -\frac{GM}{a} \left[\frac{3}{2}+\frac{\mu}{\sqrt{x^2+y^2}}+ \frac{3}{2} x^2\right]
\end{eqnarray}
Setting the derivative of $\Psi(x,0)$ equal to zero defines the symmetric Lagrange 1 and 2 points along the $x$ axis at 
\begin{eqnarray}
 r_h = \pm (\mu/3)^{1/3}
\end{eqnarray}
which is defined as the Hill's sphere radius in units of the orbital separation.  The shape of the critical potential surface is then given by setting the potential equal to its value at $x=r_h$ and $y=0$:
\begin{eqnarray}
Y^2 =\frac{4/9}{(1-X^2/3)^2}-X^2
\end{eqnarray}
 where $X$ and $Y$ are $x$ and $y$ scaled to $r_h$.
 
 A plot of this critical potential surface is shown in Fig.~\ref{fig:Roche} with the L1 and L2 points labeled.  
  
We now wish to examine what orbit is followed by a small chunk of the asteroid (i.e., hereafter a ``fragment'') which breaks off at the L1 point, begins to fall toward the host star, and is then pushed into orbit via coriolis forces.  We do this by computing the energy and angular momentum in inertial space of a test particle that is released from rest at the L1 point in the rotating frame:
\begin{eqnarray}
\mathcal{E} & ~=~ & -\frac{2}{1-r_h}-\frac{2\mu}{r_h}+ (1-r_h)^2 \\
\mathcal{J} & ~=~ & (1-r_h)^2 -3r_h^2
\end{eqnarray}
where $\mathcal{E}$ is the specific energy in units of $GM/(2a)$, and $\mathcal{J}$ is the specific angular momentum in units of $\sqrt{GMa}$.  The first  term in the expression for $\mathcal{J}$ is the initial angular momentum of the fragment, while the second term is the angular momentum that is lost by the fragment when its orbit is torqued by the asteroid during its escape.  The dependence of the second term on $r_h^2$ can be shown analytically, but we determined the leading factor of 3 by direct numerical integration.

\begin{figure}[h!]
\begin{center}
\includegraphics[width=0.5 \textwidth]{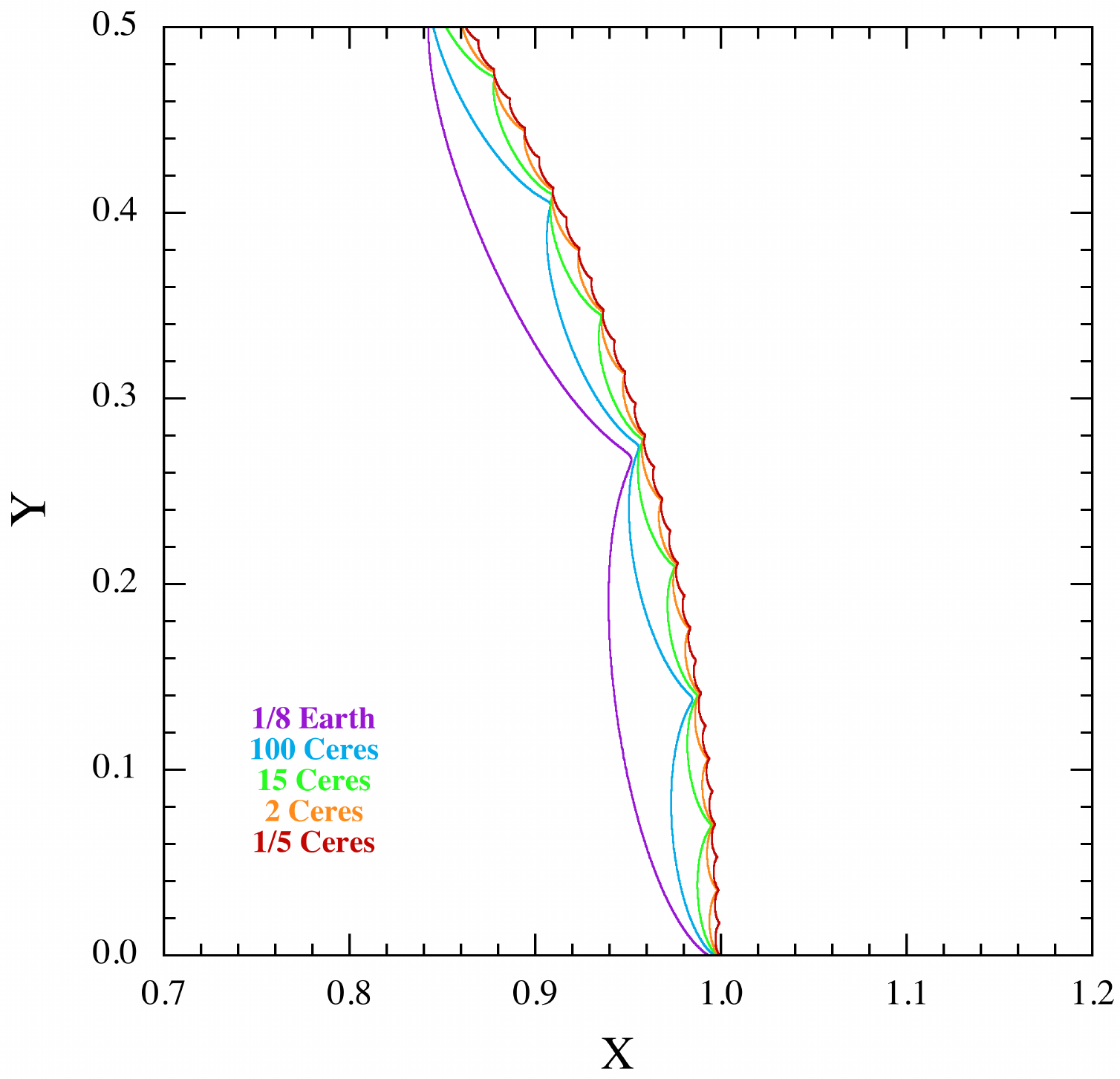}
\caption{Illustrative orbits of fragments that are released from rest at the L1 point of the parent asteroid.  These tracks were numerically integrated from their release point.  $X$ and $Y$ are in units of the semimajor axis of the parent asteroid.  For each track, the mass of the asteroid, $M_a$, is decreased by a factor of 8; they cover the range from 1/8th of an Earth mass to 1/5 the mass of Ceres.  Because the Hill's sphere radius scales as $M_a^{1/3}$, the eccentricity and drift period scale as $r_{\rm h}$, and therefore vary in steps of a factor of 2.}
\label{fig:orbits}
\end{center}
\end{figure}

\begin{figure}[h!]
\begin{center}
\includegraphics[width=0.5 \textwidth]{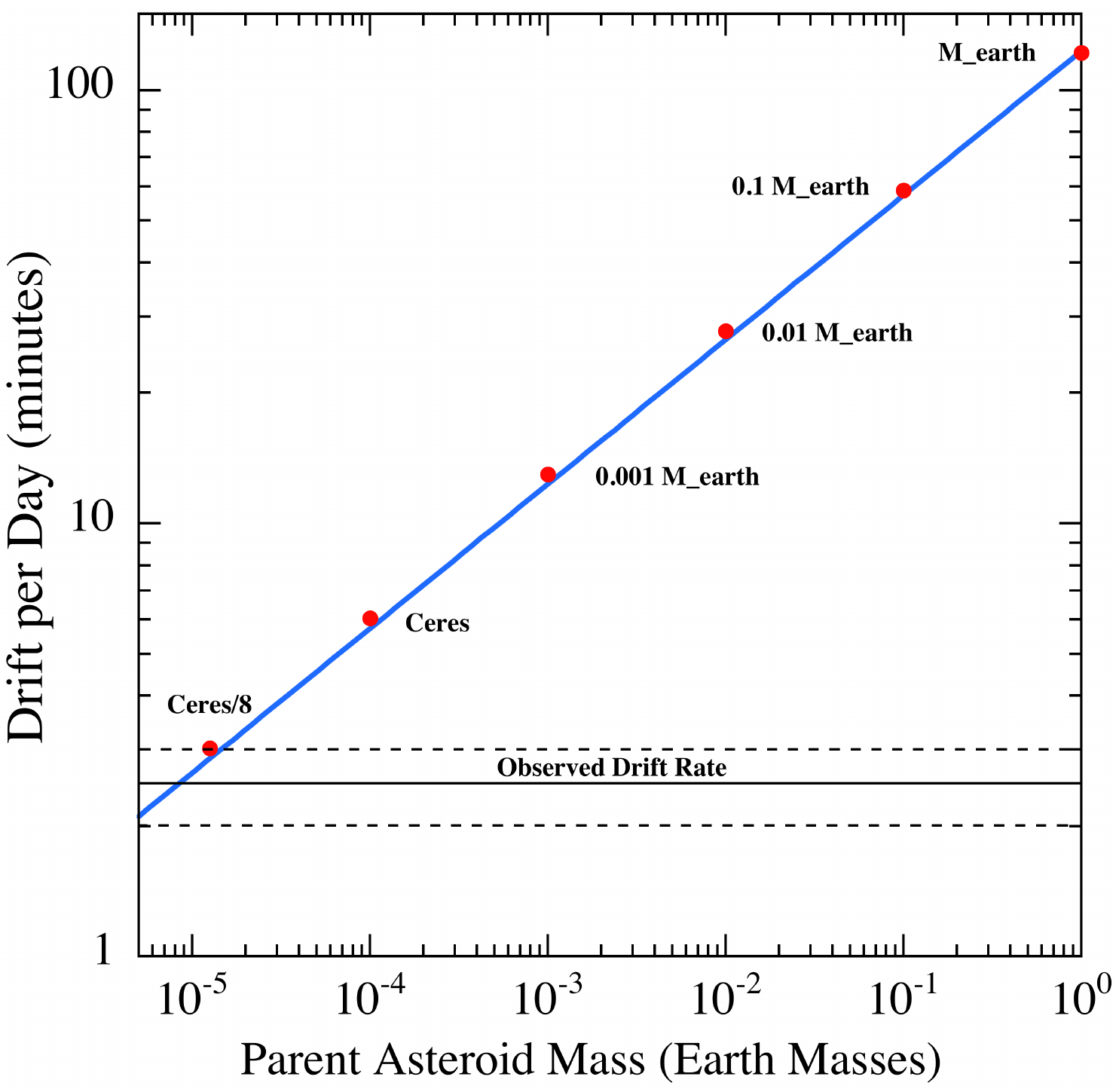}
\caption{Drift rate of broken-off fragments with respect to the asteroid as a function of the mass of the asteroid.  See Eqn.~(\ref{fig:drift}).  The mass of the asteroid associated with the ``A'' period is inferred to be about 1/10 the mass of Ceres.  The red points are based on numerical integration, while the blue curve is a plot of Eqn.~(\ref{eqn:driftP})}
\label{fig:masses}
\end{center}
\end{figure}

The orbital eccentricity is given by:
\begin{eqnarray}
e = \sqrt{1+ \mathcal{J}^2 \mathcal{E}} \simeq 3 r_h
\end{eqnarray}
while the semimajor axis of the asteroid is
\begin{eqnarray}
a' = a/|\mathcal{E}| \simeq a(1- 4 r_h)
\end{eqnarray}
The orbital period of the fragment is given by
\begin{eqnarray}
P_f = P_a/|\mathcal{E}|^{3/2} \simeq P_a (1-6 r_h)
\end{eqnarray}

We show portions of five illustrative tracks for the released fragments in Fig.~\ref{fig:orbits}.  The different tracks are for five different masses of the asteroid from which the fragment originates (varying in mass steps by factors of 8).  From this, one can visualize how both the orbital eccentricity and drift rate of the fragment depend on $r_h \propto M_a^{1/3}$, and both indeed scale as $\mu^{1/3}$. 

The drift period with which the fragment departs from the asteroid in the rest frame of the asteroid has an analytic solution:
\begin{eqnarray}
\frac{P_a-P_f}{P_a} \simeq 6 r_h = 2 \times 3^{2/3} \mu^{1/3}
\label{eqn:driftP}
\end{eqnarray}
If the mean drift rate that we observe is $2.5 \pm 0.5$ minutes per day, this implies that 
\begin{eqnarray}
\frac{P_a-P_f}{P_a} \simeq  0.0017 \pm 0.0004 \simeq 2 \times 3^{2/3} \mu^{1/3}
\label{fig:drift}
\end{eqnarray}
We can then solve for $\mu$ to find $(7 \pm 3) \times 10^{-11}$.  If we take the white dwarf mass to be $0.6 \, M_\odot$ then, the mass of the asteroid is $M_p \simeq (8 \pm 4) \times 10^{22}$ g, or about 1/10 times the mass of Ceres.  The general results for the mass determination of the asteroid in terms of the drift velocity of the daughter fragments are summarized in Fig.~\ref{fig:masses}. 

We can also use the above value of $\mu$ to estimate the physical size of the Hill's sphere, namely: $R_{\rm h} \simeq a\, (\mu/3)^{1/3} \approx 225$ km.  In our scenario, this also provides a first estimate for the size of the asteroid itself.

\end{document}